\documentclass[twocolumn,notitlepage,prx,superscriptaddress,citeautoscript,showpacs,amsmath,amstex,amssymb,mathfonts,showpacks]{revtex4-1}
\pdfoutput=1
\usepackage{amsthm,color,amsfonts,graphicx,verbatim}
\usepackage{amsmath}
\usepackage{amssymb}
\usepackage{amsthm}
\usepackage{amsfonts}
\usepackage{listings}
\usepackage{enumerate}
\usepackage{latexsym}
\usepackage{bm}
\usepackage{graphicx}
\usepackage{color}
\usepackage{rotating}
\usepackage{float}
\usepackage{hyperref}
\hypersetup{
    bookmarks=true,         
    unicode=false,          
    pdftoolbar=true,        
    pdfmenubar=true,        
    pdffitwindow=false,     
    pdfstartview={FitH},    
    pdftitle={Superconductivity and Nematic Fluctuations in a model of doped FeSe monolayers: A Determinant Quantum Monte Carlo Study},    
    pdfauthor={P. T. Dumitrescu et. al.},     
    pdfsubject={},   
    pdfcreator={},   
    pdfproducer={}, 
    pdfnewwindow=true,      
    colorlinks=true,       
    linkcolor=black,          
    citecolor=blue,        
    filecolor=magenta,      
    urlcolor=blue           
}

\newcommand{\corr}[1]{\langle{ #1}\rangle}
\newcommand{\bi}{\mathbf{i}}
\newcommand{\bj}{\mathbf{j}}
\newcommand{\bq}{\mathbf{q}}
\newcommand{\br}{\mathbf{r}}
\newcommand{\ud}{\mathrm{d}}

\DeclareMathOperator{\Tr}{Tr}
\DeclareMathOperator{\Det}{Det}
\newcommand{\1}{xz}
\newcommand{\2}{yz}

\usepackage{color}
\newcommand{\note}[1]{{\color{black}{#1}}}
\newcommand{\noteM}[1]{{\color{black}{#1}}}

\begin{document}

\begin{titlepage}

\author{Philipp T.~Dumitrescu}
\email{philippd@berkeley.edu}
\author{Maksym Serbyn}
\affiliation{Department of Physics, University of California, Berkeley, CA 94720, USA}
\author{Richard T.~Scalettar}
\affiliation{Department of Physics, University of California, Davis, CA 95616, USA}
\author{Ashvin Vishwanath}
\affiliation{Department of Physics, University of California, Berkeley, CA 94720, USA}
\affiliation{Materials Science Division, Lawrence Berkeley National Laboratories, Berkeley, CA 94720, USA}

\title{Superconductivity and Nematic Fluctuations in a model of doped FeSe monolayers: A Determinant Quantum Monte Carlo Study
}
\date{\today}

\begin{abstract}
{In contrast to bulk FeSe, which exhibits nematic order and low temperature superconductivity, \note{highly doped FeSe} reverses the situation, having high temperature superconductivity appearing alongside a suppression of nematic order.  To investigate this phenomenon, we study a minimal electronic model of FeSe, with interactions that enhance nematic fluctuations. This model is sign problem free, and is simulated using determinant quantum Monte Carlo (DQMC). We developed a DQMC algorithm with parallel tempering, which proves to be an efficient source of global updates and allows us to access the region of strong interactions. Over a wide range of intermediate couplings, we observe  superconductivity with an extended $s$-wave order parameter, along with enhanced, but short-ranged, $q=(0,0)$ ferro-orbital (nematic) order. These results are consistent with  approximate weak-coupling treatments that predict that nematic fluctuations lead to superconducting pairing. Surprisingly, in the parameter range under study, we do not observe nematic long-range order. Instead, at stronger coupling an unusual insulating phase with $q=(\pi,\pi)$ {\em antiferro}-orbital order appears, which is missed by weak-coupling approximations. }
\end{abstract}
\pacs{05.30.Rt,74.25.Dw,74.70.Xa,74.40.Kb}
\maketitle
\end{titlepage}

\section{Introduction}
A remarkable recent development in materials science has been the observation of enhanced superconductivity in single layers of FeSe, grown initially on SrTiO$_{3}$ (STO) substrates~\cite{Kun:2012,Hu:2012}. In contrast to bulk FeSe which undergoes a superconducting transition at a  relatively low temperature $T_c \sim 6$~K~\cite{Hsu:2008}, $T_c$ in monolayers on STO is at least an order of magnitude larger, in excess of 60~K \cite{He:2013, Zhang:2015aa} with even higher transition temperatures reported by an unconventional transport measurement~\cite{Ge:2015aa}. Initial studies attributed the enhancement of superconductivity to coupling between electrons in the FeSe layer and an STO phonon, which was also implicated in creating shadow electron bands observed in angle resolved photoemission experiments~\cite{ZX:2014,Feng:2014}.  However, such shadow bands are also observed for electrons on the surface of STO itself, which does not superconduct~\cite{Chen:2015}. 

Many studies have observed enhancement of $T_c \sim 40$~K in FeSe in the absence of STO substrate -- for example, \note{when grown in thin layers on MgO substrate \cite{Shiogai:2015aa},} by surface electron doping by depositing potassium~\cite{Feng:2015b, Miyata:2015, Tang:2015}, or in the layered material (Li$_{0.8}$Fe$_{0.2}$)OHFeSe~\cite{Lu:2015, Sun:2015aa}. Since the phonon spectra of these materials are entirely different from STO, a \note{separate} mechanism must be at play, which is intrinsic to the FeSe layers \note{and does not depend on interface effects}. A common element between these and \note{also} the original FeSe on STO \noteM{experiments} is that heavy electron doping leads to a pair of electronlike Fermi surfaces~\cite{Miyata:2015,Tang:2015,He:2013,Tan:2013,Song:2015,ZX:2015c}. Hence we seek a mechanism for superconductivity that is {\em intrinsic} \note{to FeSe and is controlled by electron doping}.  The even higher $T_c$ of FeSe on STO is presumed to be due to a pairing boost arising from the STO phonon~\cite{DHLee:2012,ZX:2015c}, in addition to the intrinsic mechanism; \note{we will not consider this additional effect.}

What is the origin of this intrinsic $T_{c}$ enhancement? Nematic fluctuations present an appealing possibility for the following reasons: (i) Bulk FeSe undergoes a nematic transition at $100$~K, and is unique in the family of iron pnictides/chalcogenides in not having a proximate magnetic transition. In fact, no magnetic order is observed down to the lowest temperatures~\cite{Baek:2015,Baek:2015b} (ii)  Electron doping has been shown to suppress nematic order~\cite{ZX:2015c} in potassium-doped FeSe, following which superconductivity appears. (iii) Theoretically, fluctuations of nematic order in the vicinity of a nematic quantum critical point are expected to enhance superconductivity, and this effect is particularly pronounced in two dimensions~\cite{Max:2015,Scalapino:2014,Lederer:2015a}. However, existing analytical theories have focused on universal aspects of the physics and do not capture nonuniversal aspects that are relevant to experiments. On the other hand, treatments that incorporate details of FeSe band structure and interactions, often use weak-coupling or uncontrolled approximations~\cite{Scalapino:2009,Yamase:2013aa,DHLee:2012,Jiang:2015}, and may not correctly capture the true phase structure of the system. 

In this paper we investigate the role of nematic fluctuations in enhancing superconductivity, by studying a sign-problem-free model of the FeSe monolayer, using determinant quantum Monte Carlo (DQMC). The phase diagram obtained differs substantially from that predicted by the random phase approximation (RPA)~\cite{Yamase:2013aa}, particularly in the strong-coupling limit. At intermediate couplings, we find a region with substantially enhanced nematic fluctuations and superconductivity. Although there is no long-range ordered nematic, a notable feature is that the maximum enhancement of uniform nematic fluctuations coincides with the peak in a superconducting dome. Moreover, we find that superconductivity responds to doping in an essentially asymmetric way -- electron doping enhances, while hole doping suppresses superconductivity.  All these findings link the emergence of superconductivity  to nematic fluctuations, and are  potentially relevant for the physics of FeSe films.

Other models, which were recently studied using DQMC~\cite{Berg:2012aa,Lederer:2015b}, introduce  the order parameter -- such as antiferromagnetic or nematic order -- as a separate, dynamic bosonic degree of freedom. Moreover, superconductivity was not observed in the effective model considered by \cite{Lederer:2015b}. While such an approach  is appropriate to studying universal aspects of quantum phase transitions, here we will be interested in more microscopic questions. We emphasize that our model defined below includes only electronic degrees of freedom, with properly chosen interactions that are sign problem free. Similar techniques can be used to study many other multiorbital models, for which the presence of two spin species removes the sign problem at any doping.  

\section{Model} We consider a two-band tight binding model, where electrons occupy the $d_{xz}, d_{yz}$ orbitals of iron atoms on a square lattice. This simplified model captures basic features of the iron pnictide band structure~\cite{Raghu:2008aa} and allows for nematic symmetry breaking. We take the Hamiltonian

\begin{multline}\label{Eq:ham}
H = - \sum_{\bi\bj,ab, \sigma} (t^{ab}_{\bi\bj} c_{\bi a\sigma}^{\dag} c_{\bj b\sigma}^{\phantom{\dag}} + \textrm{H.c.}) - \mu \sum_{\bi, a} n_{\bi, a}
\\-\frac{g}{2} \sum_{\bi} (n_{\bi,\1}-n_{\bi,\2})^2,
\end{multline}
where $a,b = \1,\2$ are orbital indices, $\sigma = \uparrow, \downarrow$ is the spin index, and $n_{\bi,a} = \sum_{\sigma}c^\dagger_{\bi a\sigma}c^{\phantom{\dag}} _{\bi a\sigma}$ is the occupation of orbital $a$ on lattice site $\bi$. 

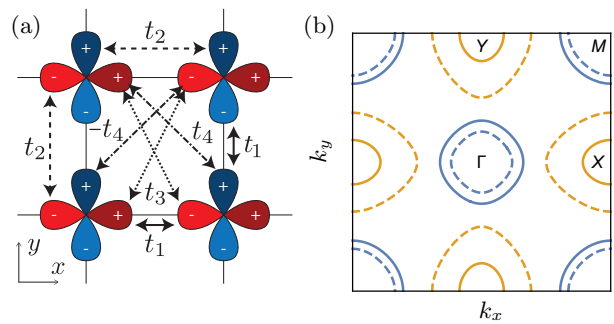
\begin{figure}[tb]
\begin{center}
\includegraphics[width=0.45\columnwidth]{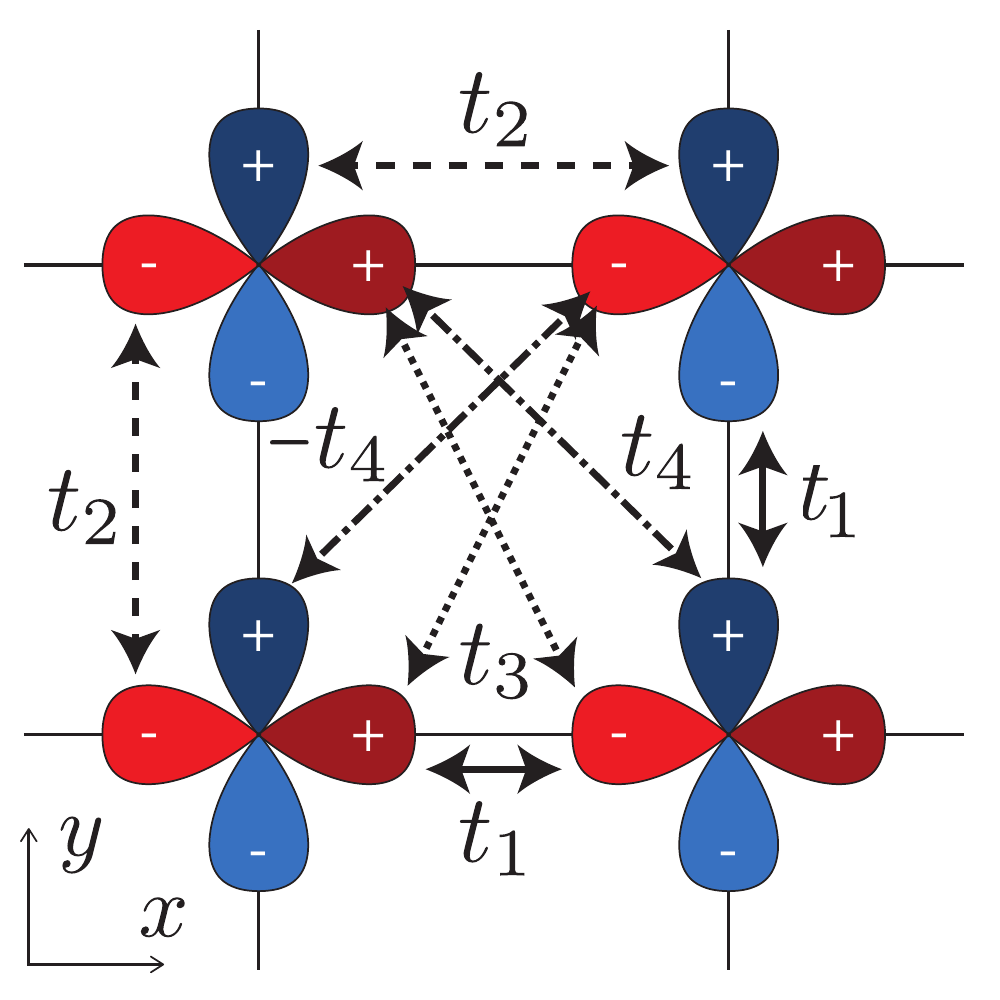}%
\qquad
\includegraphics[width=0.4\columnwidth]{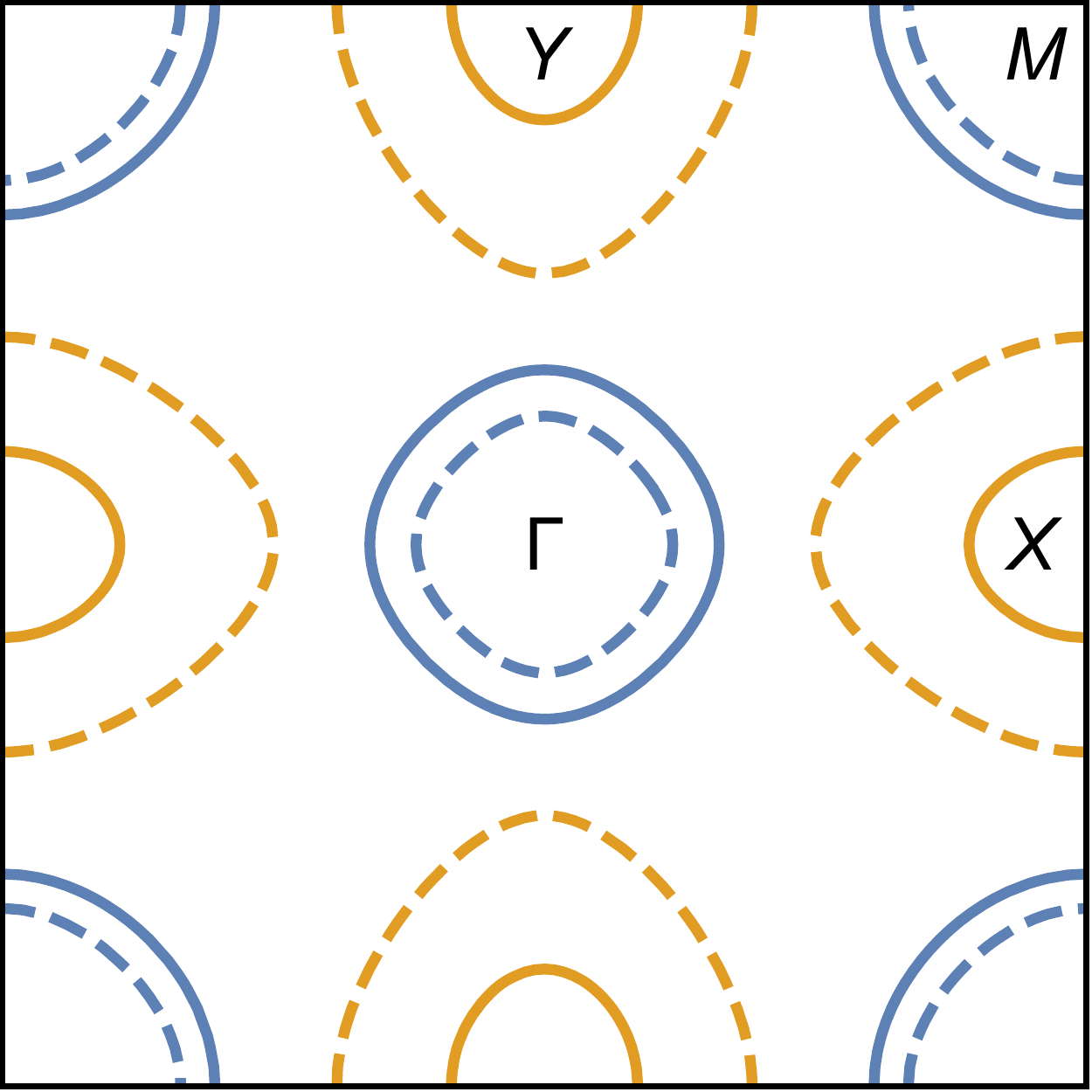} 
\setlength{\unitlength}{\linewidth}
\begin{picture}(0,0)
\put(-.95, .4){(a)}
\put(-.5, .4){(b)}
\put(-0.23, -0.039){$k_x$}
\put(-0.49, 0.2){\rotatebox{90}{$k_y$}}
\end{picture}
\end{center}
\caption{\label{Fig:1}
(a) Definitions of hopping coefficients $t^{ab}_{\bi\bj}$, where $d_{xz}$ and $d_{yz}$ are schematically shown in red and blue. All hoppings except for $t_4$ conserve orbital index.
(b) Fermi surface for chemical potential $\mu=0.6$ (solid lines) and $\mu=2$ (dashed lines) at $g=0$, showing two hole pockets at $\Gamma, M$ and two electron pockets at $X, Y$. The hopping values are $t_{1}  = -1.0,\, t_{2} = 1.5,\, t_{3} = -1.2,\, t_{4} = -0.95$.  The filling fractions {per site per orbital} are $f=0.43,\, 0.58$ at $\mu = 0.6,\, 2$ respectively. Throughout, we will work in the one iron unit-cell convention.}
\end{figure}

Allowed hopping coefficients $t^{ab}_{\bi\bj}$ are dictated by the symmetry of the $d_{xz,yz}$ orbitals and we include hopping between nearest-neighbor ($t_{1}, t_{2}$) and next-nearest-neighbor sites ($t_{3}, t_{4}$), as shown in Fig.~\ref{Fig:1}(a). The values of $t_{1,\ldots,4}$ coincide with those used in Ref.~\cite{Yamase:2013aa}; we will measure energy in units of~$t_1$. The Fermi surface in the non-interacting limit ($g=0$) with chemical potential $\mu = 0.6$ consists of two electron pockets at $X,Y$ and two hole pockets at $\Gamma, M$~[Fig.~\ref{Fig:1}(b)].  Upon increasing $\mu$ the hole pocket at $M$ disappears, while the electron pockets grow. 

The attractive interaction term ($g>0$) in the second line of Eq.~\eqref{Eq:ham} favors an on-site nematic symmetry by splitting the two orbitals and breaking $C_{4}$ rotational symmetry. This is characterized by a nonzero order parameter $\delta n_{\bi} =n_{\bi,\1}-n_{\bi,\2}$. Since the interaction is strictly on-site, the pattern of any orbital ordering is not specified a priori. 

The weak-coupling RPA considers the leading instability of the system from the free fermion susceptibility and predicts a variety of orders for Eq.~\eqref{Eq:ham}, depending on the value of $\mu$. In the range $0.2\lesssim\mu\lesssim2.5$, including the original parameters considered in \cite{Yamase:2013aa}, the RPA predicts the onset of uniform nematic order for $g_{c} \approx1.7$. For $\mu\gtrsim2.5$, the susceptibility peaks at  wavevectors $(0,\pi)$ and $(\pi,0)$ predicting stripe order, while for $\mu\lesssim0.2$ the susceptibility peaks  at wavevector $(\pi,\pi)$ predicting antiferro-orbital (AFO) (antiferroquadrupole) order. 

When interactions dominate $g\gg t,\mu$, we can get intuition from a strong-coupling expansion in $t/g$.  At zeroth order, the ground state is doubly degenerate --  either orbital $\1$ or $\2$ is fully occupied on each site. This degeneracy is split by second order processes, leading to nearest- and next-nearest-neighbor Ising-type interactions of order $\sim t^2/g$. For our hopping parameters, the ground state of the resulting Ising model is a checkerboard pattern \cite{Landau:1980, Landau:2009} -- this corresponds to AFO order at half-filling ($f=0.5$ electrons per site per orbital per spin). On the other hand, intuition from antiferromagnetic order in the half-filled Hubbard model \cite{Hirsch:1985aa} suggests that doping will quickly destroy this checkerboard order. 

The sensitivity of the weak-coupling instability to $\mu$, along with the instability of AFO order with respect to doping away from half-filling suggest a number of competing orders and we proceed to study the phase diagram of Eq.~\eqref{Eq:ham} numerically.
 
\section{Sign-problem-free DQMC} We simulate the model in an unbiased fashion using DQMC with discretized time steps as described in \cite{Blankenbecler:1981aa,White:1989aa,Assaad:2008aa}. In order to sample the full imaginary time partition function $Z = \Tr e^{-\beta H} $, we  decouple the interactions in the nematic channel using an on-site, continuous Hubbard-Stratonovich field $\varphi$; the interaction term is $\sim \sum_\bi \varphi_\bi \delta n_\bi$. Integrating out the fermions analytically, the partition function becomes a path integral of the auxiliary field, $Z = \int {D}{\varphi} \ e^{-S_{b}(\varphi)} \Det G^{-1}(\varphi)$, and can be sampled using the Monte Carlo algorithm. The fermion determinant  $\Det G^{-1}(\varphi)$ decouples into  spin sectors since the kinetic energy does not mix spins and $\varphi_\bi$ couples equally to $\uparrow, \downarrow$ through $\delta n_{\bi}$. The spin sectors are equal by time reversal for any field configuration $\varphi$, $\Det G^{-1}(\varphi) =  \Det G^{-1}_\uparrow(\varphi)\Det G^{-1}_\downarrow(\varphi)= |\Det G^{-1}_\uparrow(\varphi)|^2>0$, which guarantees that the partition function can be sampled in a sign-problem-free manner at any filling.

We perform sweeps through the space-time lattice and update the Hubbard-Stratonovich field $\varphi$ on each site. As $\varphi$ couples different orbitals, we perform rank-two Woodbury updates \cite{Press:2007aa} when calculating $G_{\uparrow}$ on a given time-slice. We use the one-sided Jacobi singular-value decomposition algorithm~\cite{Bai:2011aa} for numerical stabilization~\cite{Loh:1989aa} on every second time slice. In order to reduce ergodicity problems at strong interactions, we run the DQMC simulation in parallel for various interaction strengths $g$ and use a parallel-tempering algorithm~\cite{Katzgraber:2006aa}, which proposes to exchange $\varphi$ configurations between simulations at different $g$ after each sweep. For the data presented here, we have simulated systems with periodic boundary conditions up to $L^{2} = 10 \times 10$ in spatial size ($200$ orbitals) with an inverse temperature of up to $\beta = 8$ ($\beta E_{F} \sim 40$); the imaginary time step is $\Delta \tau = 1/16$.

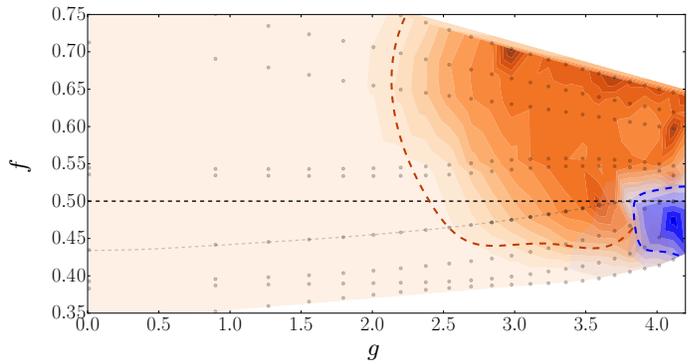
\begin{figure}[t]
\begin{center}
\setlength{\unitlength}{\linewidth}
\begin{picture}(0,0)
\put(0.0, -0.55){$g$}
\put(-0.55, -0.27){\rotatebox{90}{$f$}}
\end{picture}
\includegraphics[width=0.99\columnwidth]{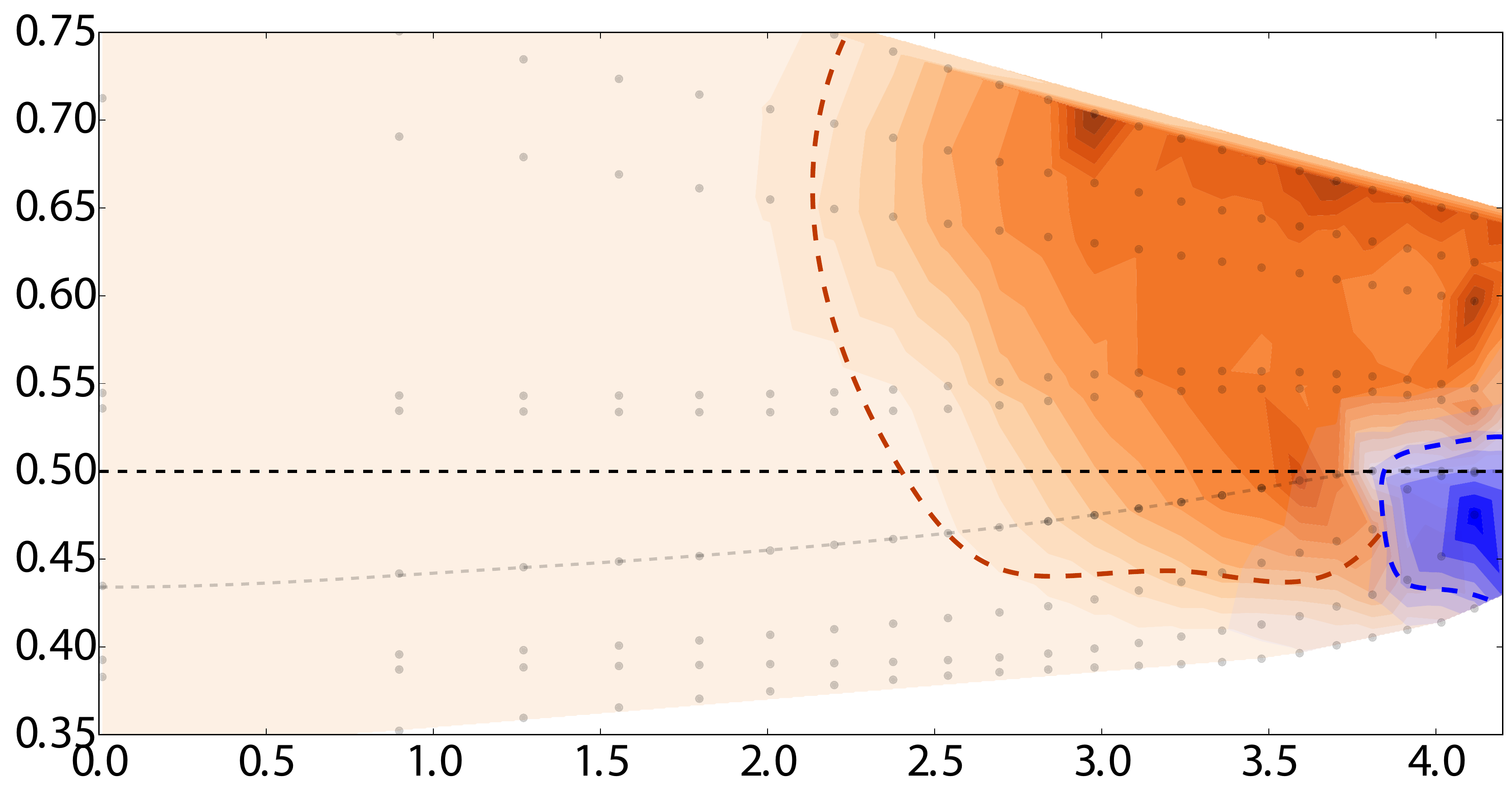} 
\end{center}
\caption{\label{Fig:2} Phase diagram of the model  Eq.~\eqref{Eq:ham} as a function of interaction $g$ and filling fraction $f$ at inverse temperature $\beta = 8$. The red dashed line indicates the boundary of the region where the superconducting order parameter extrapolates to a finite value in the thermodynamic limit. The blue dashed line marks the boundary of phase with antiferro-orbital order. The red (blue) coloring is the interpolated equal-time correlation function of the $s$-wave superconducting (antiferro-orbital) order on a $10\times 10$ lattice; white space is outside of the range sampled. Dots indicate simulated points along nine values of the chemical potentials $\mu =-1.0, -0.5, 0.0, 0.6, 1.6, 1.8, 3.0, 3.5, 4.0$; the dots joined by a gray dotted line correspond to $\mu = 0.6$. The black dashed line marks half-filling.
}
\end{figure}

\section{Phase diagram} We swept the phase diagram of the model described by Eq.~\eqref{Eq:ham}  as a function of interaction strength $g$ and filling fraction $f$ (Fig.~\ref{Fig:2}), showing regions of superconducting and antiferro-orbital order. Since we are considering a finite temperature system in two spatial dimensions, only quasi-long-range order exists. Our simulations are on lattice sizes smaller than the scale of these fluctuations and our finite-size extrapolations indicate long-range order of the $T=0$ ground state.

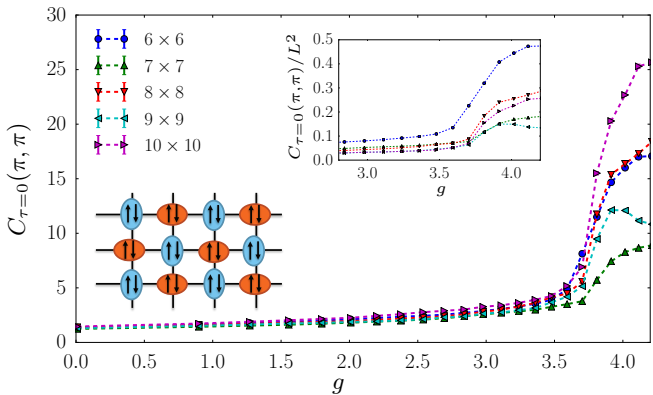
\begin{figure}[tb]
\begin{center}
\includegraphics[width=0.95\columnwidth]{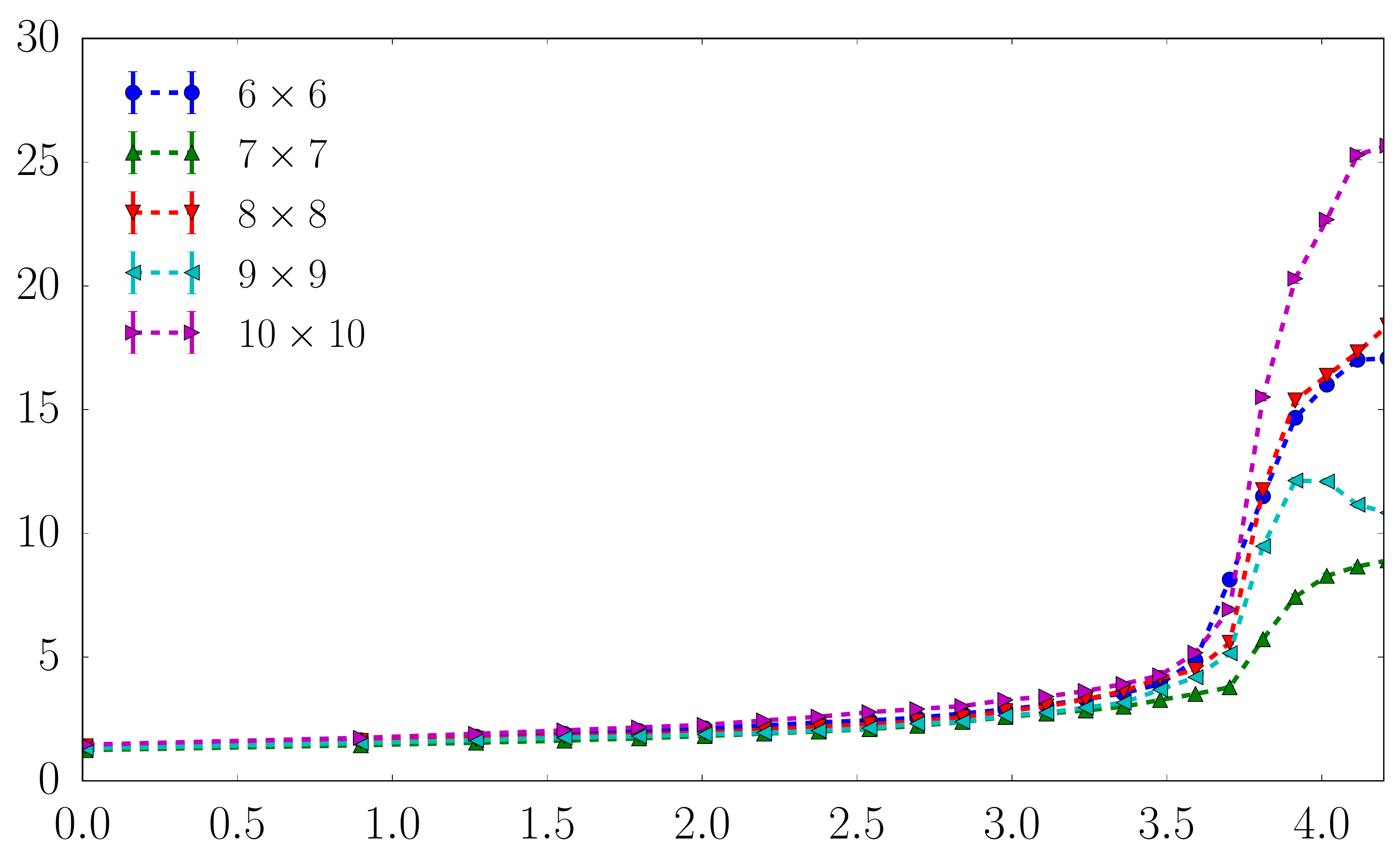}%
\setlength{\unitlength}{\linewidth}%
\begin{picture}(0,0)
\put(-0.87, 0.10){\includegraphics[width=0.3\columnwidth]{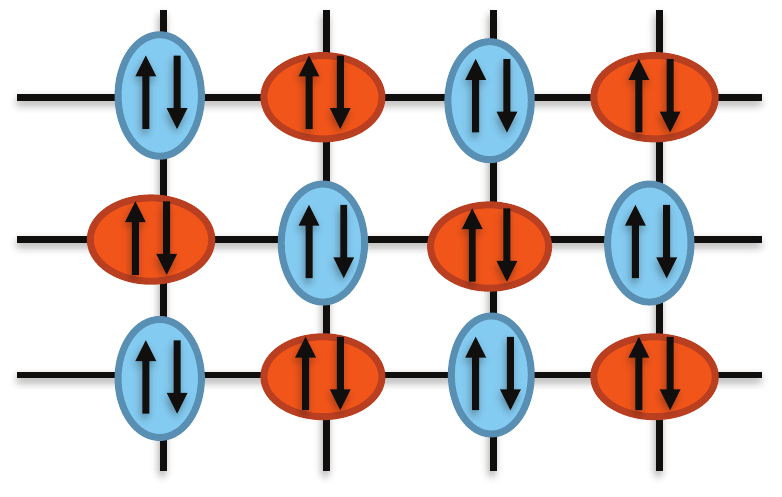}}
\put(-0.54, 0.3){\includegraphics[width=0.4\columnwidth]{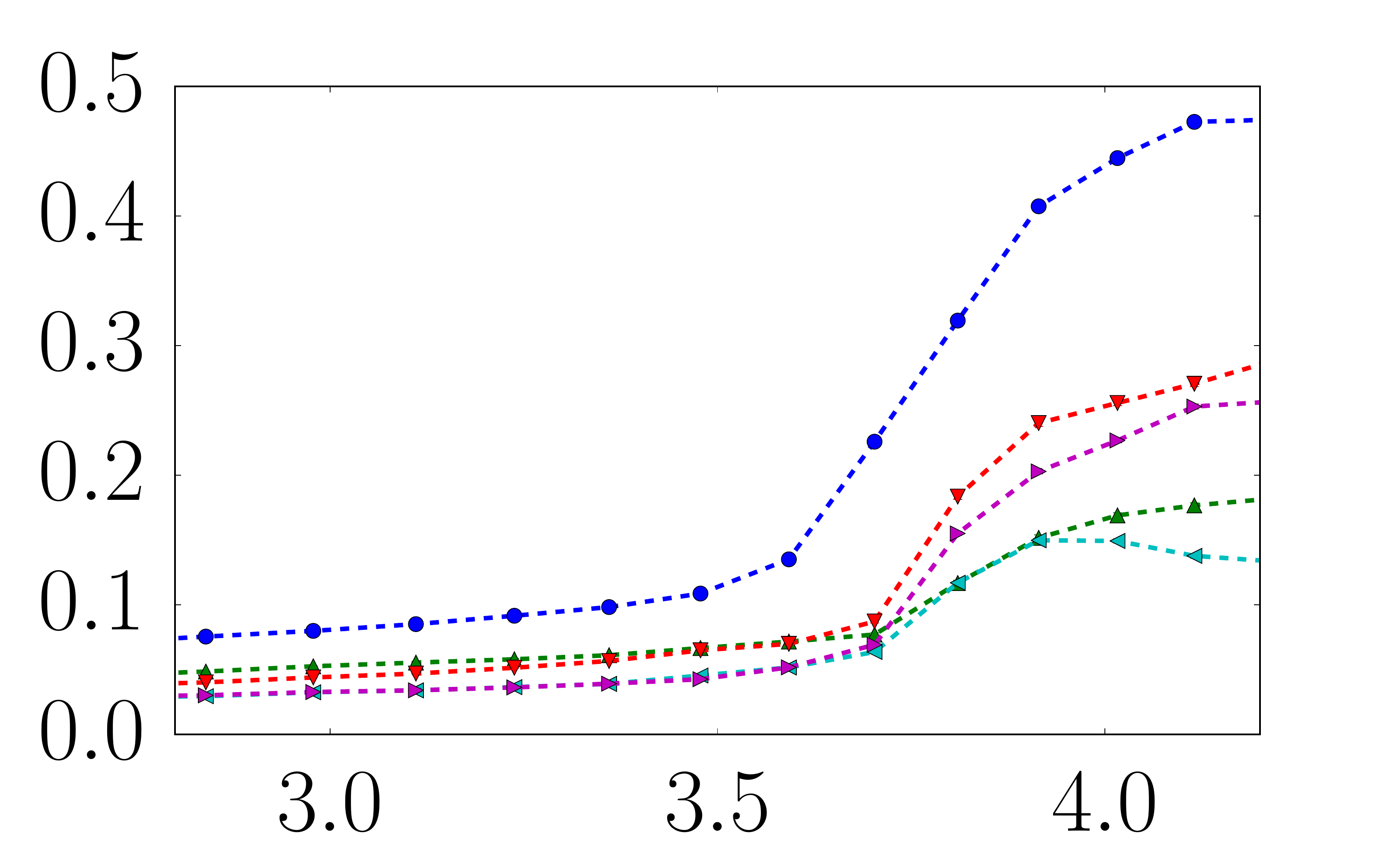}}
\put(-0.5, -0.015){$g$}
\put(-1.0, 0.22){\rotatebox{90}{$C_{\tau=0}(\pi,\pi)$}}
\put(-0.34, 0.29){${}_g$}
\put(-.57, 0.337){\rotatebox{90}{${}_{C_{\tau=0}(\pi,\pi)/L^2}$}}
\end{picture}
\end{center}

\caption{\label{Fig:3} Equal-time nematic correlation function averaged around $q=(\pi,\pi)$ rises rapidly, signaling the onset of the long-range order. The left inset shows a cartoon of the antiferro-orbital ordering pattern. The second inset indicates the convergence of the correlation function, when normalized by $L^2$, as expected from long-range order. Note the  even-odd effect due to periodic boundary conditions.}
\end{figure}

We first discuss the phase diagram in the vicinity of half-filling $f=0.5$, which corresponds to two electrons per site. In the limit of strong-coupling $g \gtrsim 3.7$ we see development of long-range antiferro-orbital order. This is fully consistent with the intuition from the strong-coupling expansion of a fully polarized state in the orbital basis with a checkerboard ordering pattern~(Fig.~\ref{Fig:3} inset). The onset of order is confirmed by considering the equal time nematic correlation function

\begin{equation}\label{Eq:nem}
C_{\tau=0}(\bq) = \frac{1}{L^2}\sum_{\bi,\bj}e^{i\bq\cdot (\bi-\bj)}\corr{\delta n_{\bi}\delta n_\bj}.
\end{equation} 

\noindent The behavior of $C_{\tau=0}(\bq)$ at $q=(\pi,\pi)$ is shown in Fig.~\ref{Fig:3}. To reduce finite-size effects, we show $C_{\tau=0}(\bq)$ averaged over three neighboring points  $\bq, \bq+2\pi \hat{\textbf{x}}/L, \bq+2\pi \hat{\textbf{y}}/L$ which coincide in the thermodynamic limit. We also can confirm the onset of order via the Binder ratio~\cite{Binder:1981} for the boson field $\varphi$ conjugate to $\delta n$ at zero frequency. 

The AFO order rapidly disappears when the system is doped away from half-filling, or the interaction strength is decreased. In contrast to the expectations from weak-coupling RPA, we do not observe any nematic ordering at other wave-vectors. Instead, when the long-range AFO disappears, we observe a large region with non-zero superconducting order. To probe the superconducting order, we study the equal-time pair correlation function $\sim \corr{\Delta^{\phantom{\dag}}_{ab}(\bi) \Delta_{cd}^{\dag}(\bj)}$, where the specific form of the $\Delta_{ab}(\bi)$ depends on the symmetry of pairing. We consider all possible irreducible representations of lattice point group $D_4$ involving on-site, nearest-neighbor and next-nearest-neighbor sites and found nonvanishing pair correlation function for the order parameter with $s$-wave ($A_{1}$) symmetry. The dominant response is the on-site pairing, where the only nonvanishing pairing is within the same orbitals with equal sign ($A_1\times A_1$ representation), 

\begin{equation}\label{Eq:s-ons}
\Delta^{\text{s}}(\bi) = \frac{1}{2}  c_{\bi a\alpha} (i\sigma^y_{\alpha\beta}) (\tau^0_{ab})c_{\bi b\beta}.
\end{equation}

\noindent Here $\sigma$ and $\tau$ are the Pauli matrices acting in the spin and orbital basis, and $\tau^0$ is an identity matrix. The order parameter $\Delta^{\text{s}}(\bi)$ coexists with the extended $s$-wave pairing between nearest neighbors,  $\Delta^{\text{s-ex}}(\bi)$, where the gap changes sign between orbitals ($B_1\times B_1$ representation),

\begin{equation}\label{Eq:s-ext1}
\Delta^{\text{s-ex}}(\bi) = \frac{1}{2}  \sum_{\hat{\bf{e}}}d(\hat{\bf{e}})\, c_{\bi+\hat e, a\alpha} (\sigma^y_{\alpha\beta}) (\tau^z_{ab})c_{\bi b\beta},
\end{equation}

\noindent as is reflected by the $\tau^z$ matrix. Here, the vector $\hat{\bf e}$ runs over nearest neighbors and $d(\hat{\bf{e}} )$ denotes the $d_{x^2-y^2}$-wave symmetry form-factor, $d(\pm \hat{\bf{x}}) =1$, and  $d(\pm \hat{\bf{y}}) = -1$.  For  $\mu\geq 2$, the $s$-wave pairing also extends to next-nearest-neighbor sites, along the diagonals of the square lattice. It has a $d_{xy}$-wave form factor along with the $\tau^x$ pairing in the orbital basis~($B_2\times B_2$ representation). 

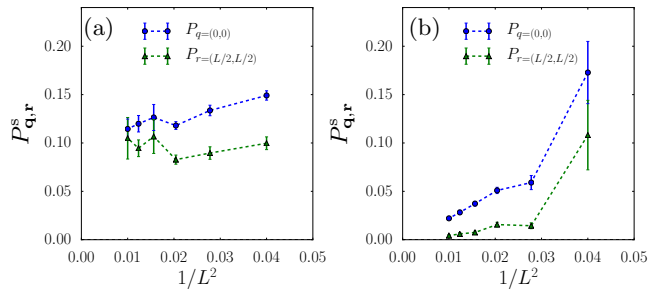
\begin{figure}
\begin{center}
\hspace{0.02\columnwidth}
\includegraphics[width=0.45\columnwidth]{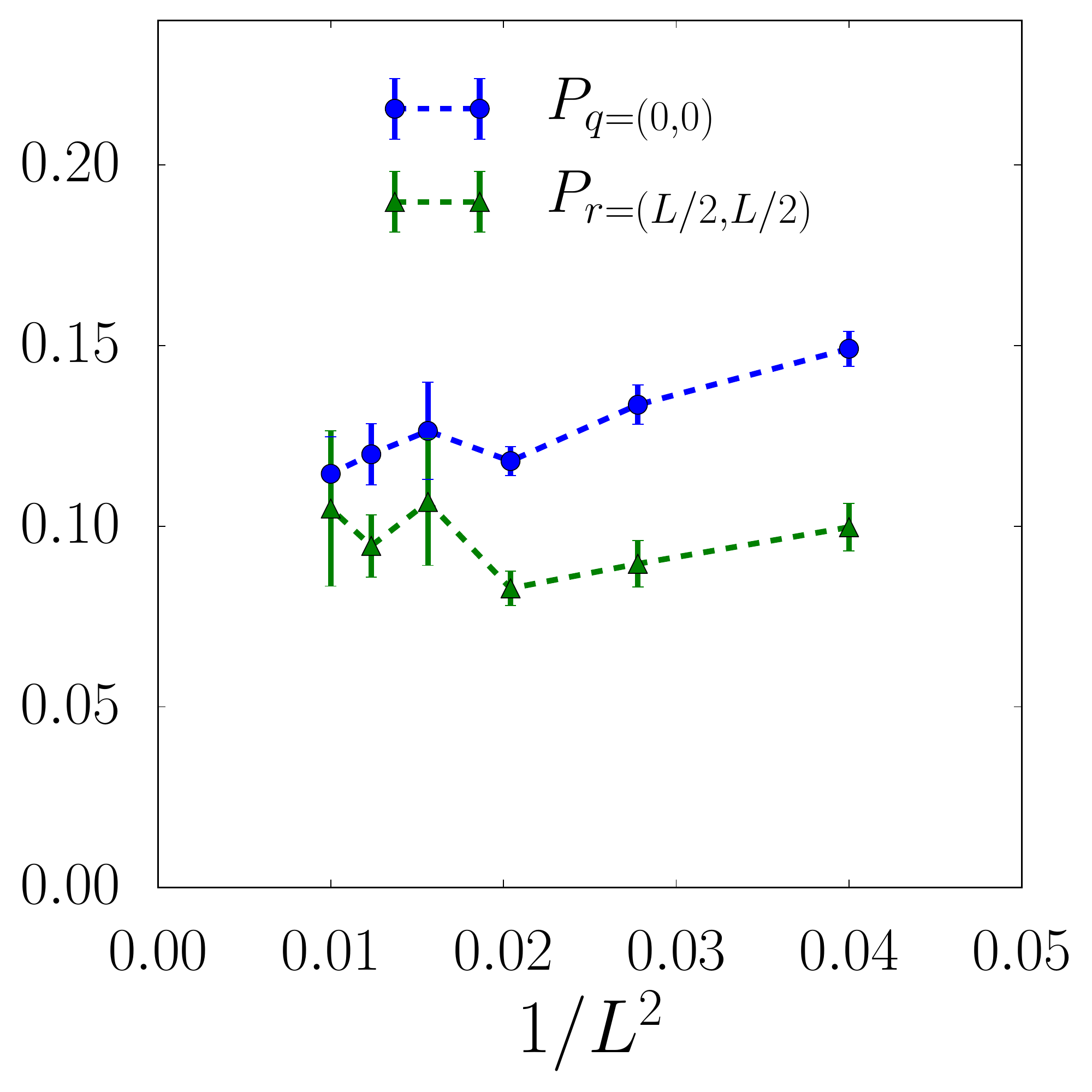} 
\hspace{0.02\columnwidth}
\includegraphics[width=0.45\columnwidth]{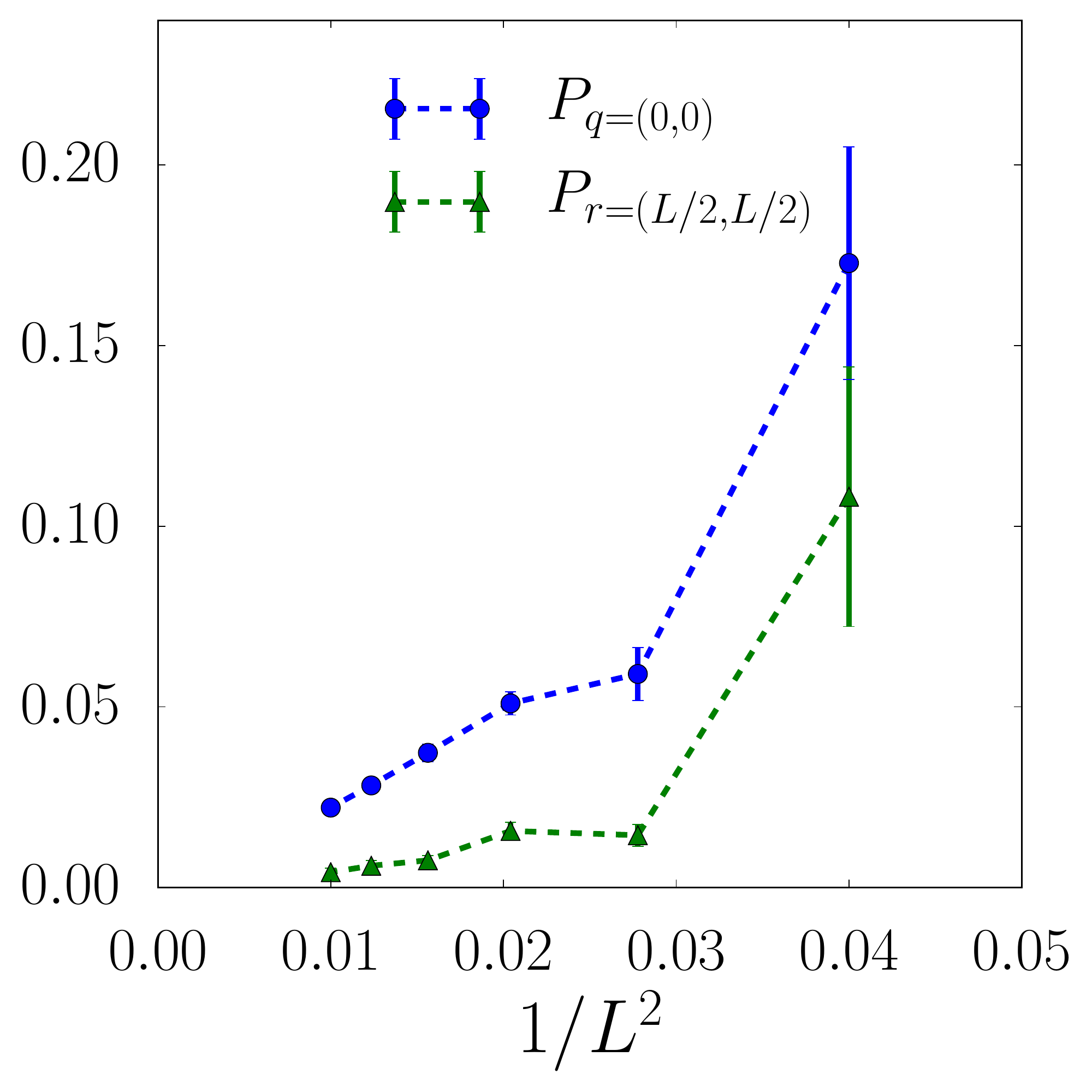}
\setlength{\unitlength}{\linewidth}
\begin{picture}(0,0)
\put(-0.90,0.40){(a)}
\put(-0.40,0.40){(b)}
\put(-1.01, 0.23){\rotatebox{90}{$P^\text{s}_{\bq,\br}$}}
\put(-0.52, 0.23){\rotatebox{90}{$P^\text{s}_{\bq,\br}$}}
\end{picture}
\end{center}
\caption{\label{Fig:4}  Finite-size scaling of the on-site $s$-wave equal-time pair correlation at maximal-distance $P^{s}_{\br = (L/2,L/2)}$ and zero momentum $P^{s}_{\bq = 0}$. (a) In the region which we identify as a superconductor ($\mu=0.6$, $g = 3.59$) $P_{\br},P_{\bq}$ extrapolate to a finite value in the thermodynamic limit.  (b) In the region of the AFO phase ($\mu=0.6$, $g=3.91$), both pair correlation functions scale to zero in the thermodynamic limit.}
\end{figure}

The equal-time $(\tau = 0)$ pair correlation function for the on-site $s$-wave is defined as 
\begin{equation}\label{Eq:Ps}
  P^{\text{s}}_\br 
  = 
  \frac{1}{L^2}\sum_\bi\corr{\Delta^{\text{s}}(\bi+\br)\Delta^{\text{s}}(\bi)},
\ 
  P^{\text{s}}_\bq
  =
  \frac{1}{L^2}\sum_\br e^{i\bq\cdot\br} P^{\text{s}}_\br
\end{equation}

\noindent in the coordinate and Fourier space, respectively, where both sums are performed over all lattice points. In the thermodynamic limit, the value of $P^{\text{s}}_\bq$ at $\bq=0$ must converge to the value of $P^{\text{s}}_\br$ at maximum separation $\br = (L/2,L/2)$, if there is long-range superconducting order. At small $L$, $P^{s}_{\bq = 0}$ includes mostly short-range contributions and overestimates the order parameter \cite{Varney:2009aa}. Figure~\ref{Fig:4}(a) shows data from the superconducting phase where both quantities extrapolate to finite value as $1/L \to 0$, moreover these quantities become closer to each other for larger system sizes. In contrast, Fig.~\ref{Fig:4}(b) shows data from the AFO phase, where the pair correlation functions are nonzero only due to finite-size effects and extrapolate to zero in the thermodynamic limit.

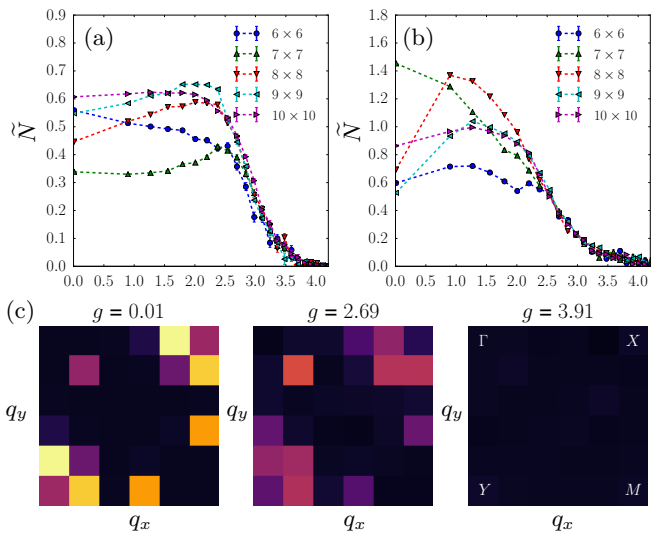
\begin{figure}[tb]
\begin{center}
\includegraphics[width=0.45\columnwidth]{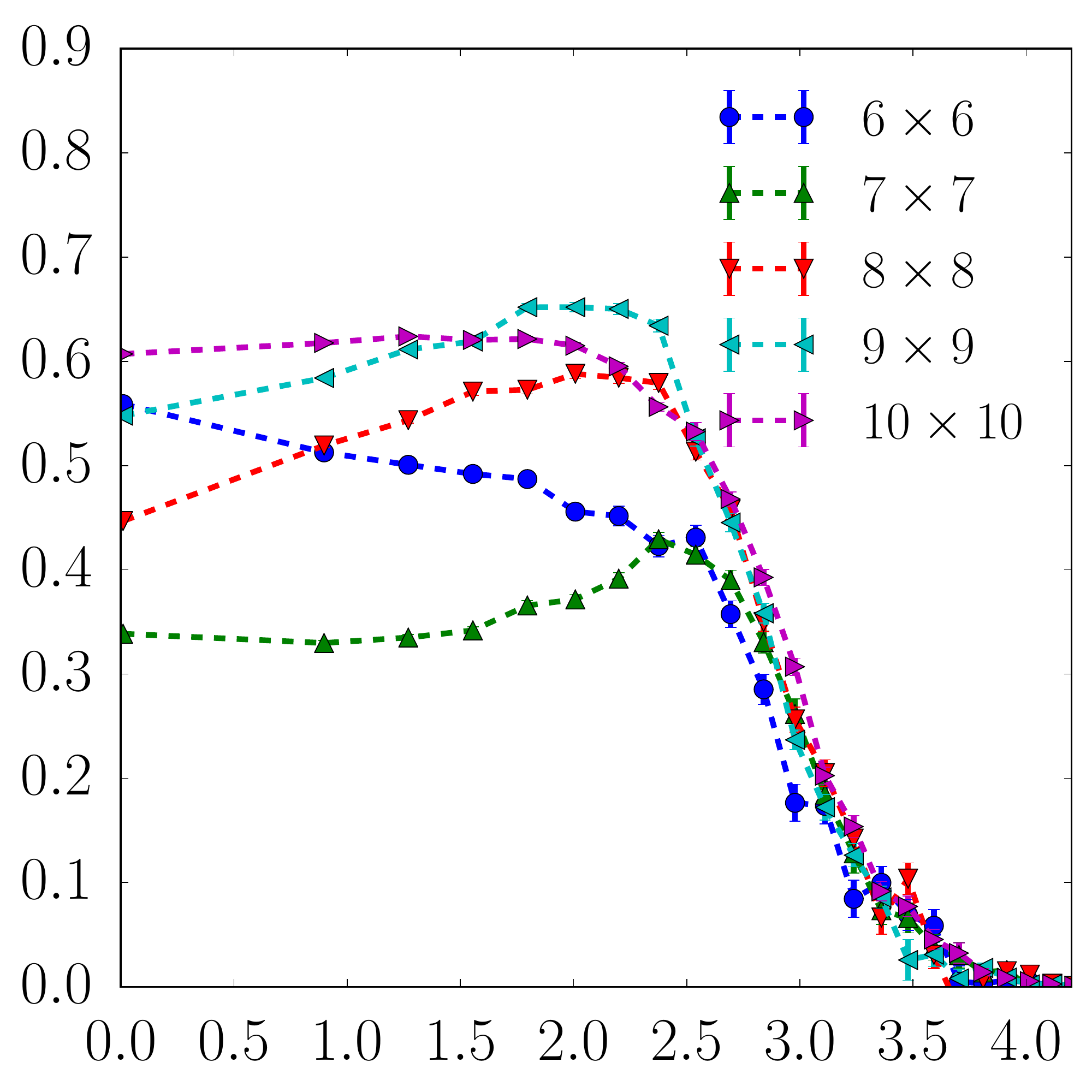} 
\hspace{0.02\columnwidth}
\includegraphics[width=0.45\columnwidth]{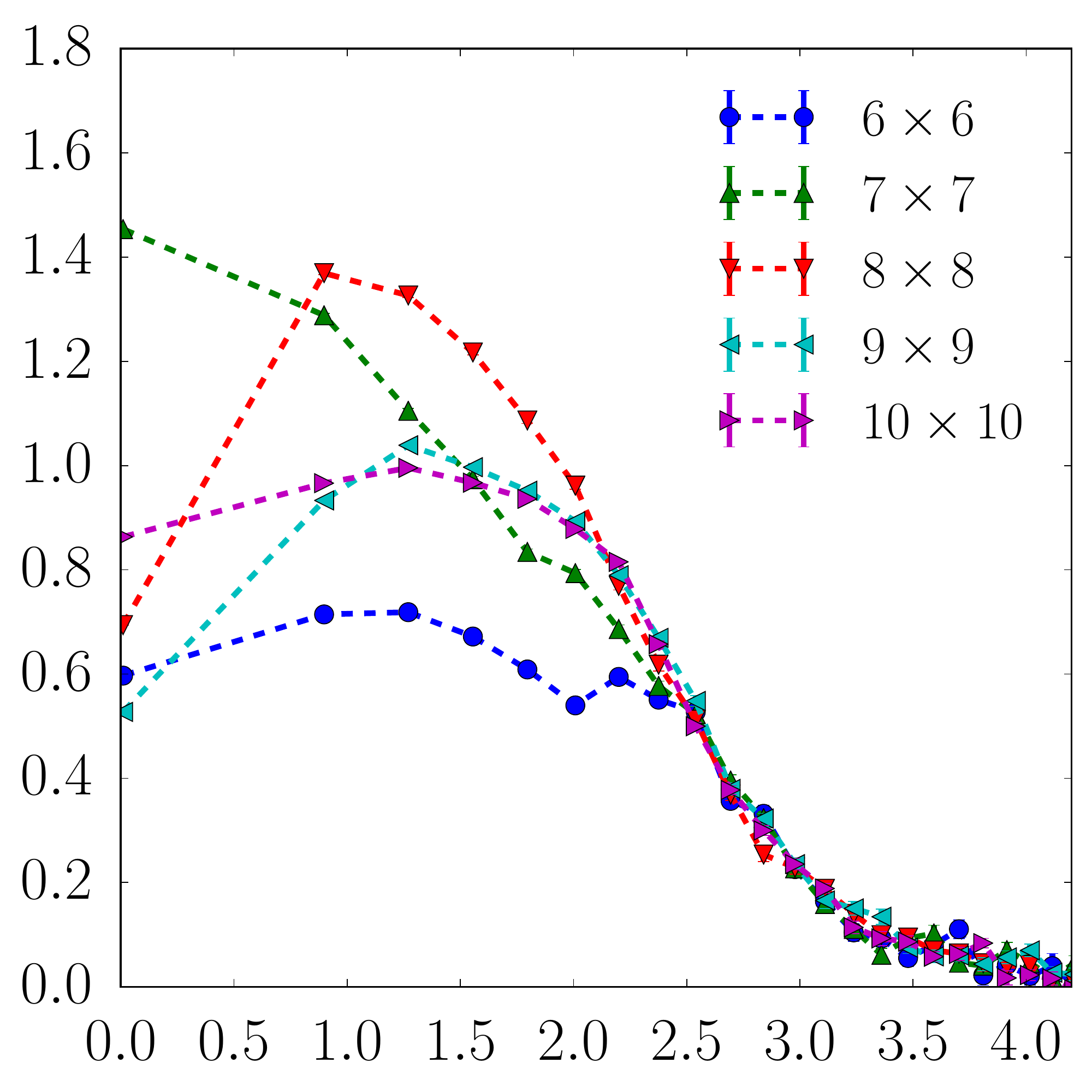} 
\vspace{20px}
\includegraphics[width=0.96\columnwidth]{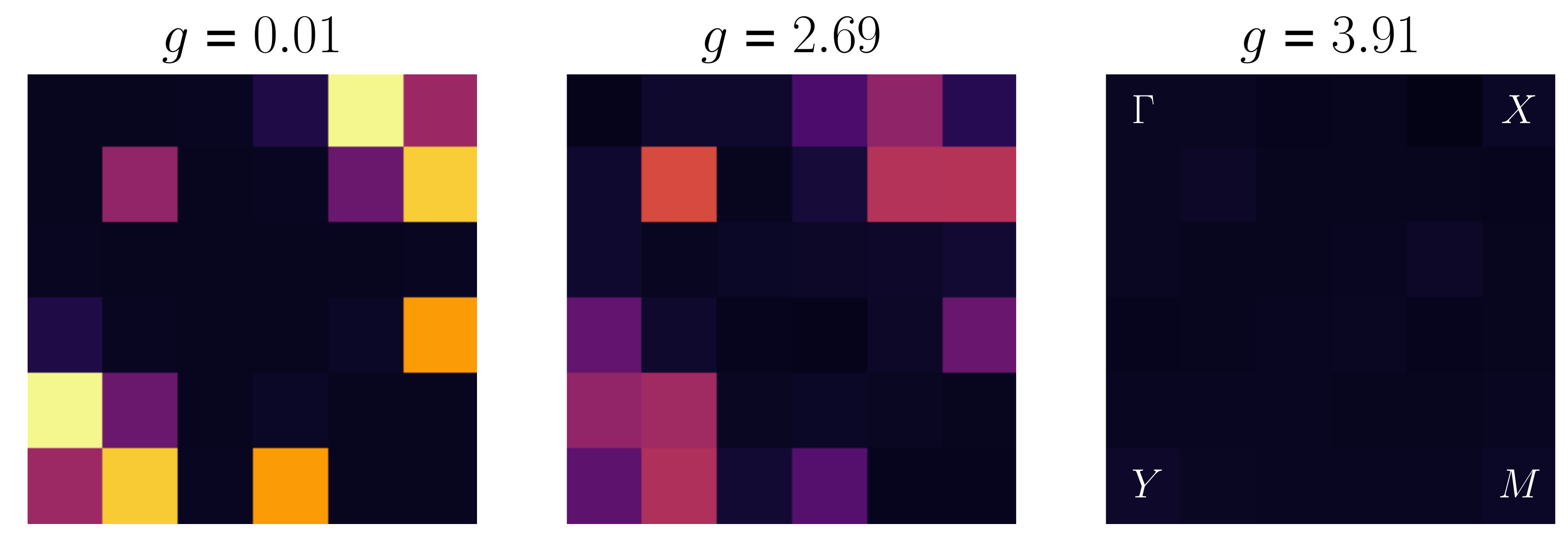} 
\setlength{\unitlength}{\linewidth}
\begin{picture}(0,0)
\put(-0.900,  0.730){(a)}
\put(-0.410,  0.730){(b)}
\put(-1.02, 0.31){(c)}
\put(-1.02, 0.16){$q_y$}
\put(-0.83, -0.01){$q_x$}
\put(-0.68, 0.16){$q_y$}
\put(-0.5, -0.01){$q_x$}
\put(-0.35, 0.16){$q_y$}
\put(-0.18, -0.01){$q_x$}
\put(-1.01, 0.58){\rotatebox{90}{$\widetilde{N}$}}
\put(-0.51, 0.58){\rotatebox{90}{$\widetilde{N}$}}
\end{picture}
\end{center}
\caption{\label{Fig:6}  The pseudo-density of states $\widetilde{N}$ along lines of constant chemical potential (a) $\mu = 0.6$ and (b) $\mu = 4.0$ shows large finite-size variation in the metallic state and starts deceasing once the system enters a gapped state.  From correlation function measurements the system in (a) enters the superconducting state at $g \sim 2.5$ and the AFO state at $g \sim 3.7$; we see here that the AFO state is gapped. The system in (b) is strongly electron doped and always far away from the AFO state; the drop to zero of $\widetilde{N}$ suggests the superconductor is fully gapped, as expected from an $s$-wave superconductor. The three bottom panels (c) show the $\bq$ space resolved pseudo-density of states of the $10 \times 10$ lattice at $\mu = 0.6$; one-quarter of the Brillouin zone is shown. As the interaction increases, the states close to the Fermi surface are gapped out.}
\end{figure}

In order to confirm the above picture of the ordered phases, we consider the pseudo-density of states \cite{Trivedi:1995aa}

\begin{equation}
\widetilde{N} =  \frac{1}{TL^2} \sum_{\bq} G\left(\tau = \frac{\beta}{2}, \bq\right) = \sum_{\bq}\int_{-\infty}^{\infty}  \frac{N(\omega,\bq)\ud \omega}{2T\cosh(\omega / 2T)}
\end{equation}

\noindent where $G(\tau,\bq)$ is the imaginary time Green's function summed over orbitals and $N(\omega,\bq)$ is the single-particle density of states at momentum $\bq$. $\widetilde{N}$ gives us a measure of the single-particle states at the Fermi energy without numerically challenging analytic continuation; in the limit where the temperature is far below any other energy scale $\widetilde{N} \simeq \pi N(\omega = 0)$.  Figure~\ref{Fig:6} shows $\widetilde{N}$ for the chemical potentials $\mu = 0.6$ (a) and $\mu = 4.0$ (b). At weak interactions, there is a finite density of states corresponding to the metallic phase with larger finite-size effects due to the discrete sampling of the Fermi surface. Once the system enters the superconducting phase, $\widetilde{N}$ drops to zero which is consistent with the fully gapped $s$-wave pairing symmetry. For (a) the system is in the AFO phase at $g \gtrsim 3.7$, which we see is insulating.


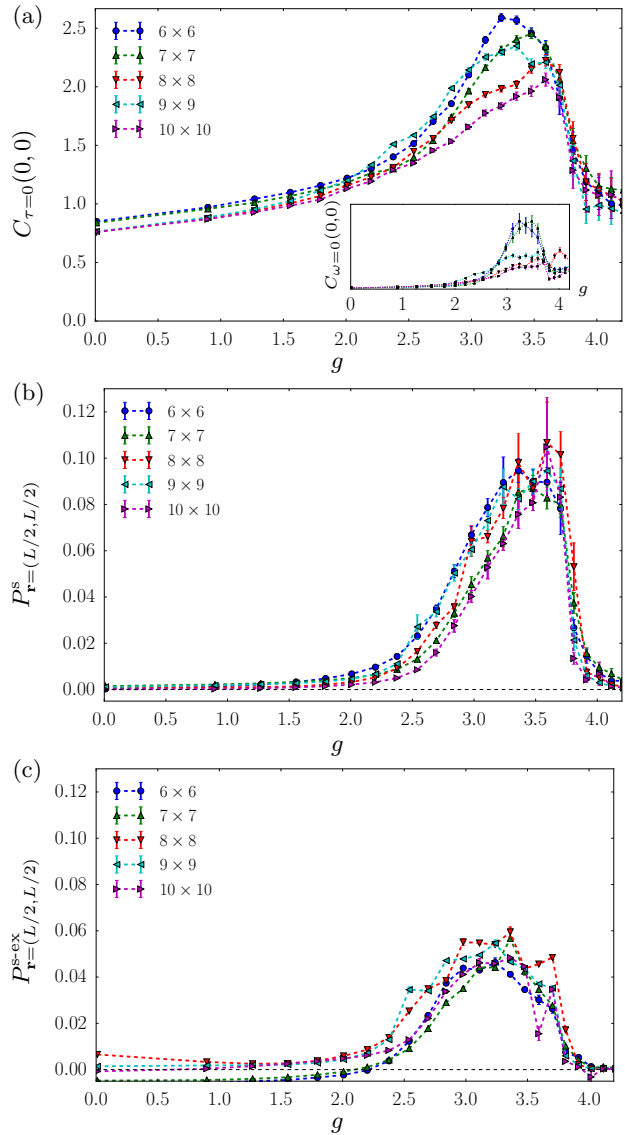
\begin{figure}
\begin{center}
\includegraphics[width=0.88\columnwidth]{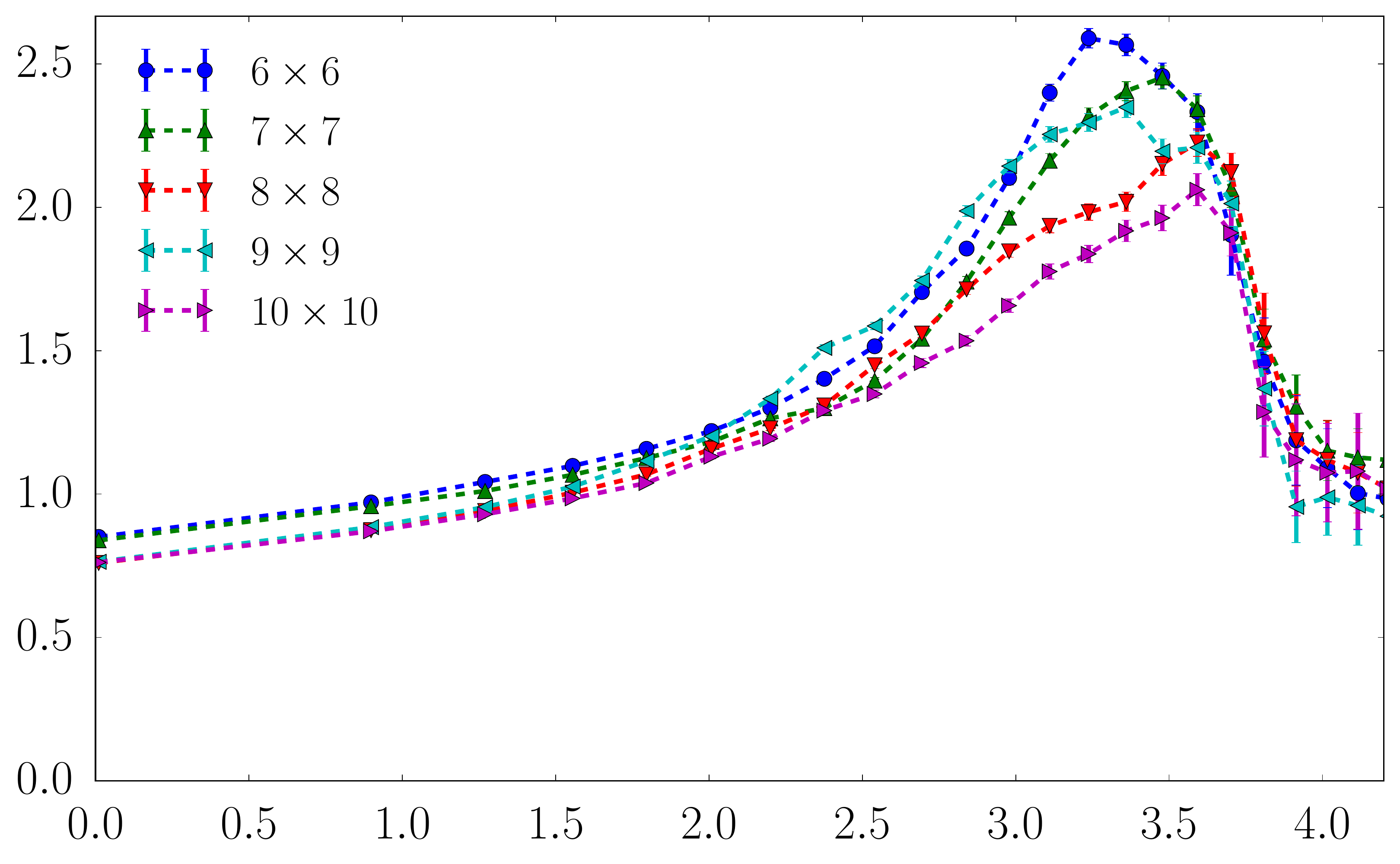}\\
\vspace{1em}
\includegraphics[width=0.88\columnwidth]{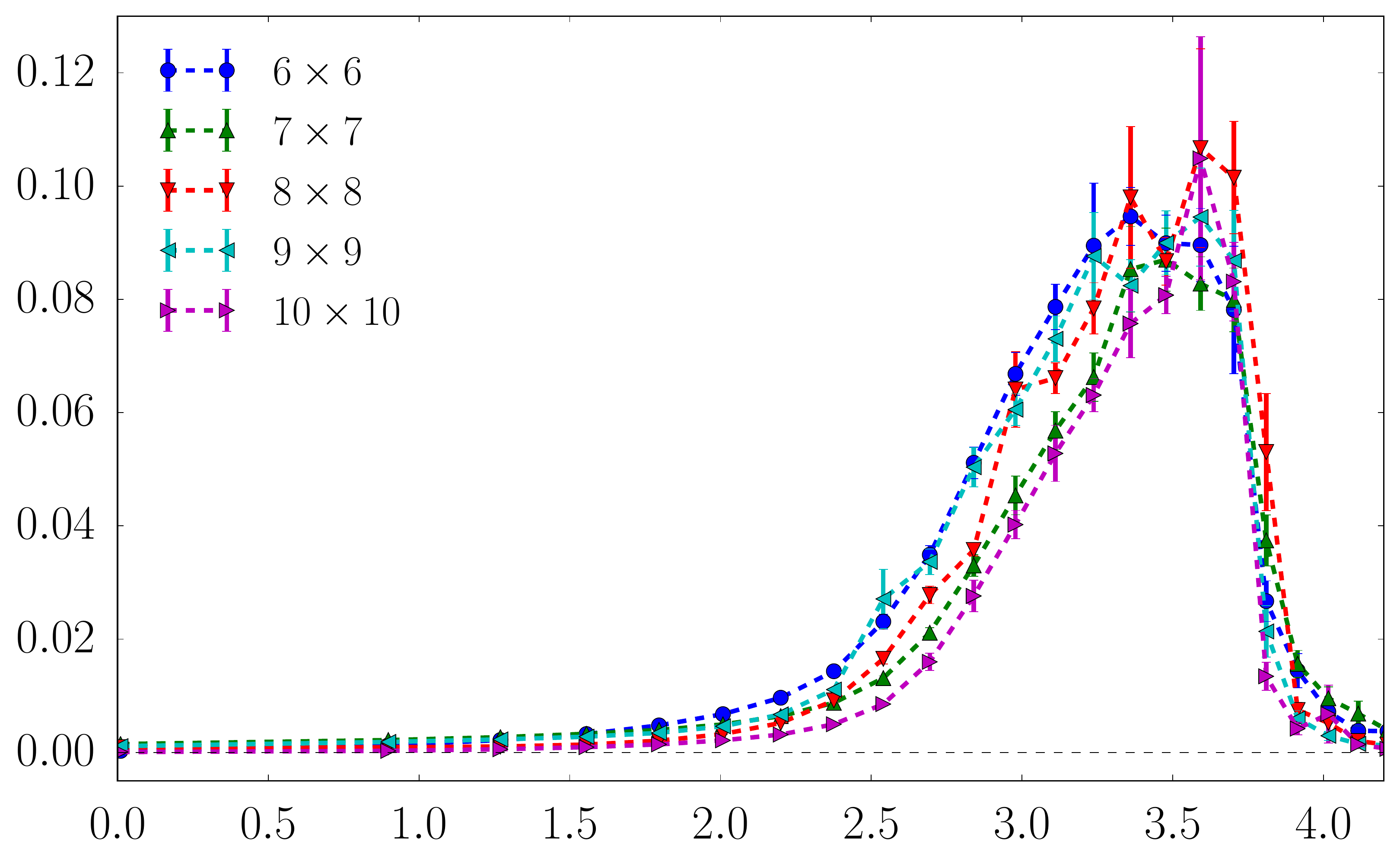}\\
\vspace{1em}
\includegraphics[width=0.88\columnwidth]{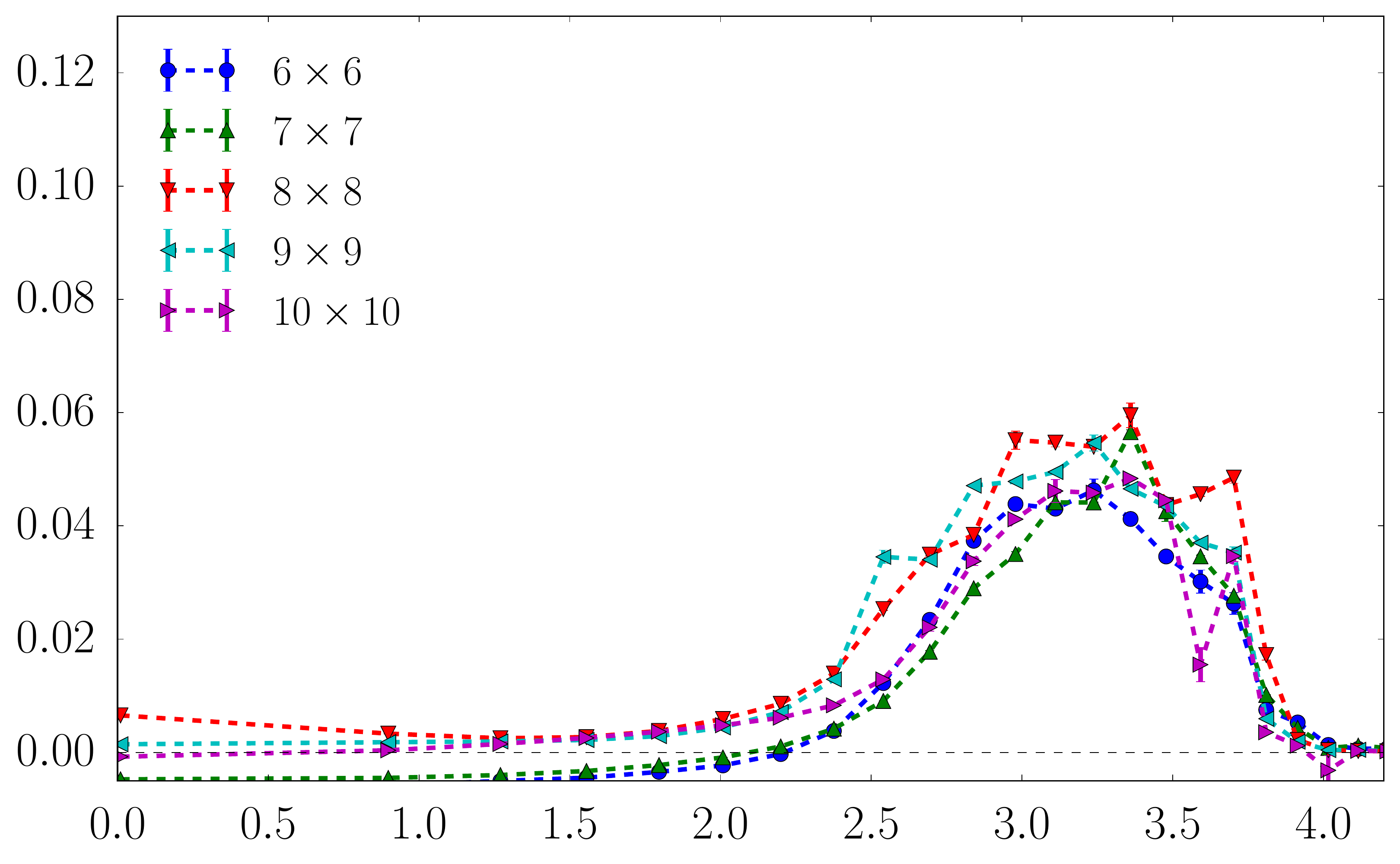}
\vspace{1em}
\setlength{\unitlength}{\linewidth}
\begin{picture}(0,0)
\put(-.45, 1.23){\includegraphics[width=0.35\columnwidth]{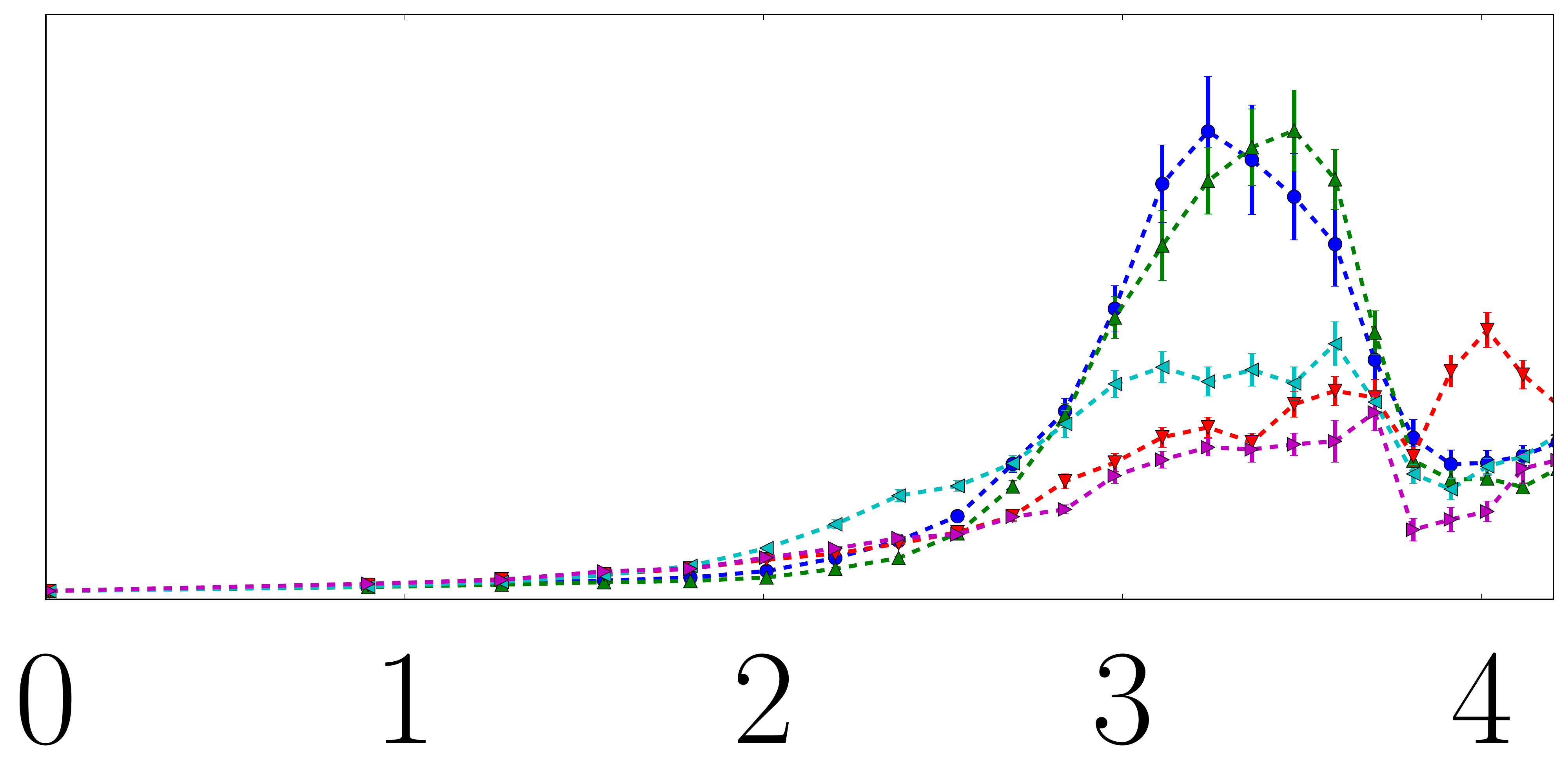}}
\put(-.09, 1.27){${}_g$}
\put(-.48, 1.27){\rotatebox{90}{${}_{C_{\omega = 0} (0,0)}$}}
\put(-0.96, 1.68){(a)}
\put(-0.96, 1.10){(b)}
\put(-0.96, 0.52){(c)}
\put(-0.96, 1.35){\rotatebox{90}{$C_{\tau=0}(0,0)$}}
\put(-0.96, .78){\rotatebox{90}{$P^\text{s}_{\br = (L/2,L/2)}$}}
\put(-0.96, .20){\rotatebox{90}{$P^\text{s-ex}_{\br = (L/2,L/2)}$}}
\put(-0.47,  1.15){$g$}
\put(-0.47,  0.56){$g$}
\put(-0.47, -0.02){$g$}
\end{picture}%
\end{center}
\caption{\label{Fig:5} The (a) uniform nematic correlation function, (b) on-site $s$-wave pair correlation function, and (c) nearest-neighbor extended $s$-wave  have very similar dependence on the interaction strength for the fixed value of $\mu=0.6$. The onset and termination of the on-site superconducting order coincides with similar trends in the nematic susceptibility. In (a), the inset shows the unnormalized nematic susceptibility $C_{\omega = 0} (0,0) = \langle \delta n \delta n\rangle$ at $q = 0, \omega = 0$ -- here we derive this data from the correlation function of the Hubbard-Stratonovitch fields, since $\varphi$ is conjugate to $\delta n$ (see also Appendix \ref{sec:app:b}).}
\end{figure}

\section{Origin of superconductivity} We observed an extended $s$-wave superconducting order in a large portion of the phase diagram.  One may worry that this superconducting order arises only from the attractive parts of the interaction in the model defined by  Eq.~\eqref{Eq:ham}. However, decoupling the interaction in the pairing channel within a mean-field calculation only leads to significant superconducting pairing for much stronger interactions, $g\geq 6$ at $\beta=8$. Since the mean-field approximation tends to overestimate the ordering tendency, this suggests that this scenario in isolation is improbable. 

 A number of recent works~\cite{Lederer:2015a, Max:2015} addressed enhancement of superconductivity in the vicinity of a uniform nematic transition by nematic fluctuations. While we do not find long-range uniform nematic order in the considered range of doping and interactions, the intuition from weak-coupling RPA suggests possible competition between various ordering tendencies for this model. Then, upon approaching the AFO transition, we expect to have enhancement of fluctuations in various channels, including uniform nematic fluctuations. 

To check if the uniform nematic fluctuations play a role in the superconducting phase, we compare the evolution of equal-time nematic and pair correlation functions with interactions, Figs.~\ref{Fig:5}(a)-(c). The uniform nematic correlation function has a maximum around $g\approx 3.5$, exactly where $P^{\text{s}}_\br$ peaks. For larger interaction, the onset of the AFO phase signaled by a rapid increase in AFO correlations for $g\geq 3.7$~(see Fig.~\ref{Fig:3}) coincides with the destruction of superconductivity and suppression of uniform nematic correlations.  

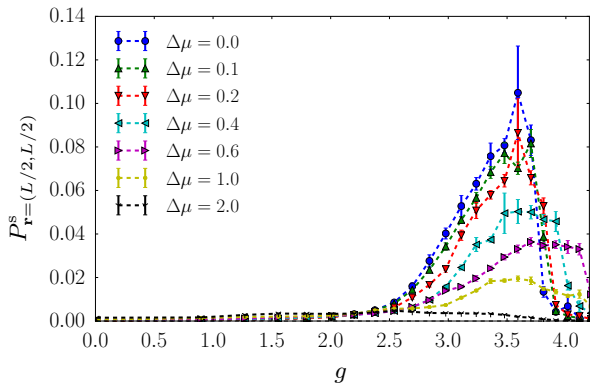
\begin{figure}[tb]
\begin{center}
\includegraphics[width=0.98\columnwidth]{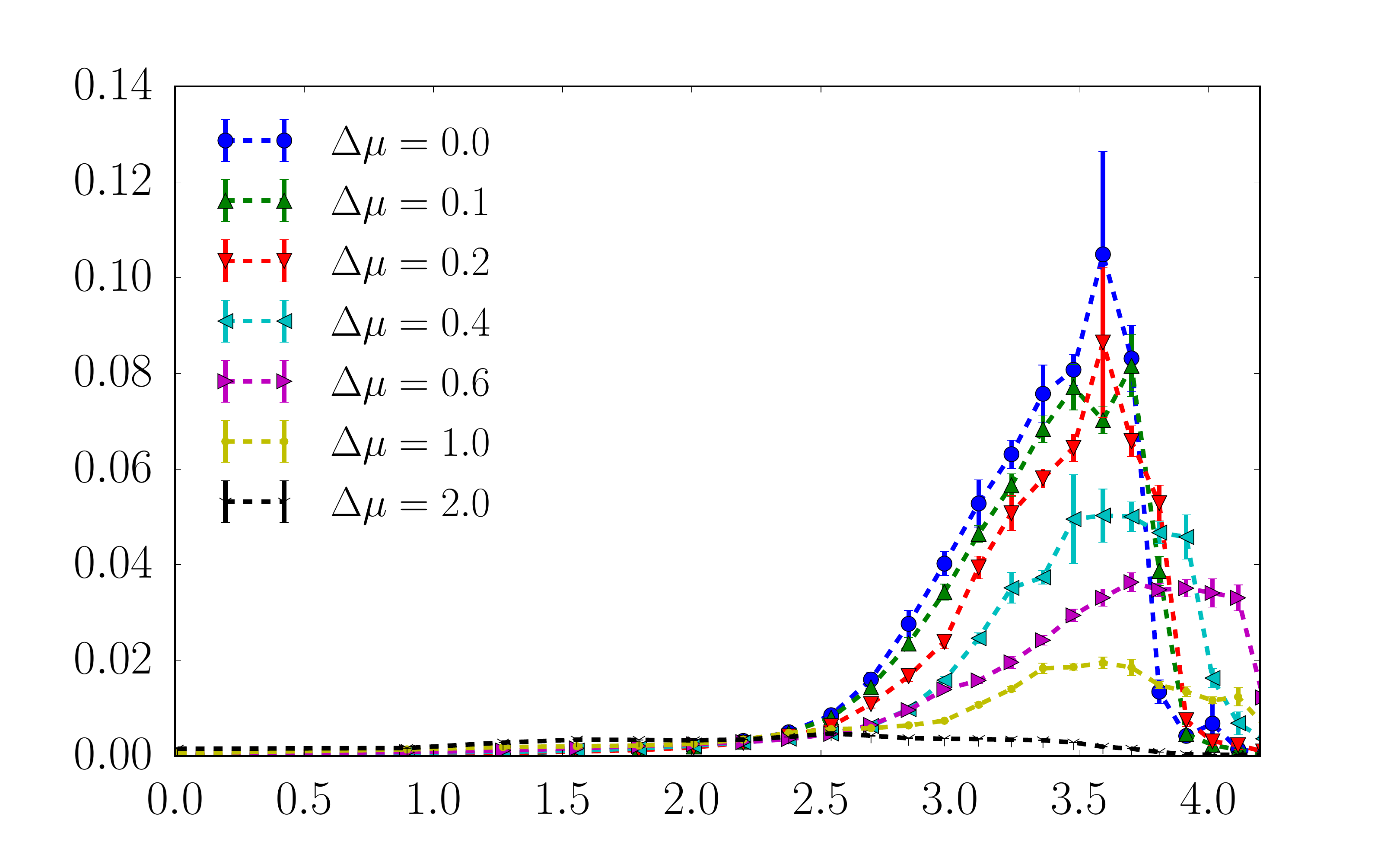}\\
\setlength{\unitlength}{\linewidth}
\begin{picture}(0,0)
\put(-0.5, .24){\rotatebox{90}{$P^\text{s}_{\br = (L/2,L/2)}$}}
\put(0, 0.03){$g$}
\end{picture}%
\end{center}
\caption{\label{Fig:7} On-site $s$-wave correlation is suppressed in simulations which have an explicit orbital symmetry breaking term $  \Delta \mu \sum_\bi  \delta n_\bi/2$. The simulations are performed on a $10 \times 10$ lattice at constant chemical potential $\mu = 0.6$, so the dependence of filling on $g$ is different from the case $\Delta\mu=0$. Note for $\Delta \mu \gtrsim 0.4$, pairing persists to higher values of the interaction strength $g$.  This is understandable since  the filling is shifted to higher electron doping where the insulating phase is suppressed.
}
\end{figure}

To further explore the relationship between uniform nematic fluctuations and superconductivity, we consider  adding an explicit symmetry breaking term  $ \Delta \mu \sum_\bi  \delta n_\bi/2$ to the Hamiltonian. This suppresses the uniform nematic fluctuations by causing the system to order in one of the orbitals; the superconducting order (Fig.~\ref{Fig:7}) is strongly suppressed with increased symmetry breaking. However, the symmetry breaking term also causes a change in the band structure and pushes the filling dependence to higher electron doping (at a fixed chemical potential $\mu$), which we also expect to modify the superconducting response. While it is hard to isolate the impact of these different effects, the following comparison may be worthwhile -- an orbital splitting of $\mu =0.6$ leads to a 65\% suppression of superconductivity at $g=3.0$. A uniform chemical potential change of the same magnitude starting at the same filling leads to a suppression of 67\% for hole doping and a 60\% enhancement for electron doping. When we split the orbitals, we expect the effect from changing the band structure to be in between the cases of uniform electron/hole doping and have a small effect. Since the suppression is strong, the orbital splitting appears to have a more significant impact on superconducting pairing which we attribute to the suppression of the nematic fluctuations.

\section{Discussion and implications for {F\lowercase{e}S\lowercase{e}}}
Motivated by the idea of a nematic instability driven by electron-electron interactions~\cite{Fernandes:2014aa}, we considered a purely electronic model with interactions in the nematic channel.  Our studies revealed a phase diagram with a large superconducting region. While our two-band model is oversimplified, it roughly captures the behavior of the FeSe Fermi surface with doping: electron pockets increase in size upon doping, while hole pockets shrink. Moreover, we use a local on-site interaction that favors imbalance in orbital occupancy.  We consider our interaction term as an approximation after one integrates out high energy bosonic modes; similar interaction terms were shown to arise from the Fe-ion oscillations~\cite{Onari:2010}. The on-site Coulomb repulsion, which is absent in our model, will  suppress on-site pairing, but the extended parts of the superconducting pairing, which we find share the same trends as the on-site pairing, will presumably be less affected, and may be directly relevant for the observed superconductivity in FeSe films. 

Our model in Eq.~\eqref{Eq:ham} was found to have a long-range antiferro-orbital order at strong-coupling, whereas the bulk FeSe is believed to have a uniform nematic order. Nevertheless, the model considered here  has enhanced uniform nematic fluctuations as a precursor to the onset of uniform order. We found that these nematic fluctuations are correlated with enhancement of superconductivity.  Moreover, we observed an essential asymmetry of the superconducting phase: doping with electrons enhances superconducting order, while hole doping destroys it. This is consistent with the phenomenology of FeSe, where superconductivity emerges upon strong electron doping.  One can potentially try to connect the nematic fluctuation mechanism more closely with the observed superconductivity by looking for anisotropy of the gaps in momentum space~\cite{Lederer:2015a}, which is left for future work on larger system sizes. Beyond considerations of FeSe, many systems, such as intermetallic rare-earth compounds, show ferroquadrupole or antiferroquadrupole order  without any magnetic phase transitions \cite{Luthi:2007aa}. They have recently received renewed attention \cite{Rosenberg:2016aa} and our results might be relevant to the physics of these materials. 

To conclude, we proposed a two-band model with interactions which enhance nematic fluctuations and studied this model using DQMC. We find that robust high temperature superconductivity appears that is accompanied by  ferronematic fluctuations, although the ferronematic ordered phase itself does not appear in the range that was studied.  Our findings can be relevant to enhanced superconductivity in FeSe films, as well as other situations where a fluctuating order may be responsible for superconductivity. Our methods are readily extendible to a wide class of multiorbital models.

\emph{Note added.}
During completion of this work, references  \cite{Li:2015aa,Li:2015ab,Schattner:2015aa} appeared, which find superconductivity  on applying DQMC to different models. In contrast to those works which couple electrons to bosonic (typically spin density waves) modes with intrinsically defined  dynamics, here we are concerned with a purely electronic model and focus on nematic fluctuations.  Chubukov and Xing \cite{Chubukov:2016aa} recently revisit RPA based predictions whose qualitative conclusions are in broad agreement with our results.

\begin{acknowledgments}
We thank S.~Gazit for numerous discussions.  This research was supported by the Gordon and Betty Moore Foundation EPiQS Initiative through Grant No. GBMF4307 (M.S.), the U.S. Department of Energy, Office of Science, Office of Basic Energy Sciences under Grant No. DE-SC0014671 (R.T.S.), and a Simons Investigator grant (A.V.).  
\end{acknowledgments}

\appendix

{\color{black}
\section{Temperature Dependence}

\begin{figure}[tb]
	\begin{center}
		\includegraphics[width=0.95\columnwidth]{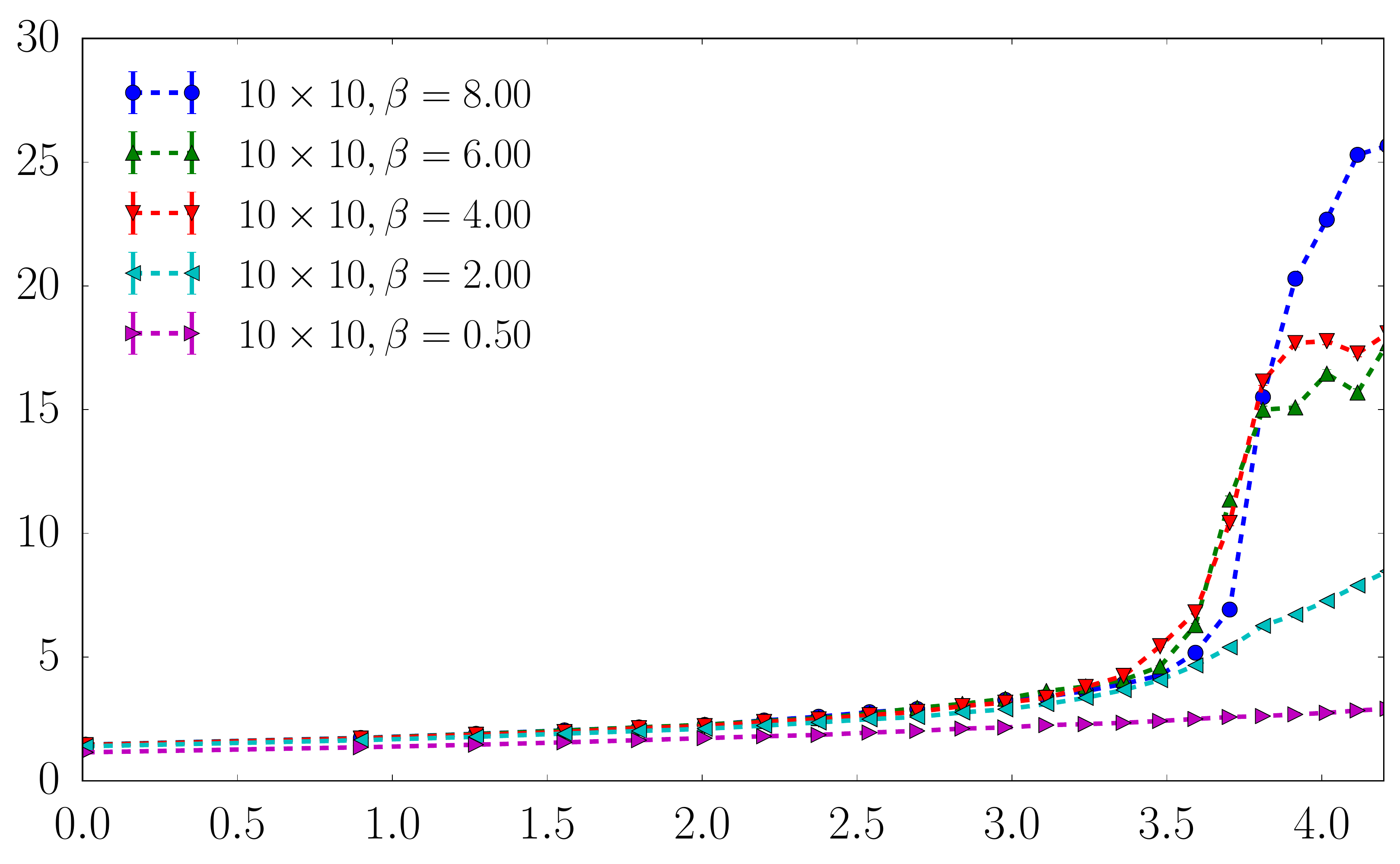}\\
		\vspace{1em}
		\includegraphics[width=0.95\columnwidth]{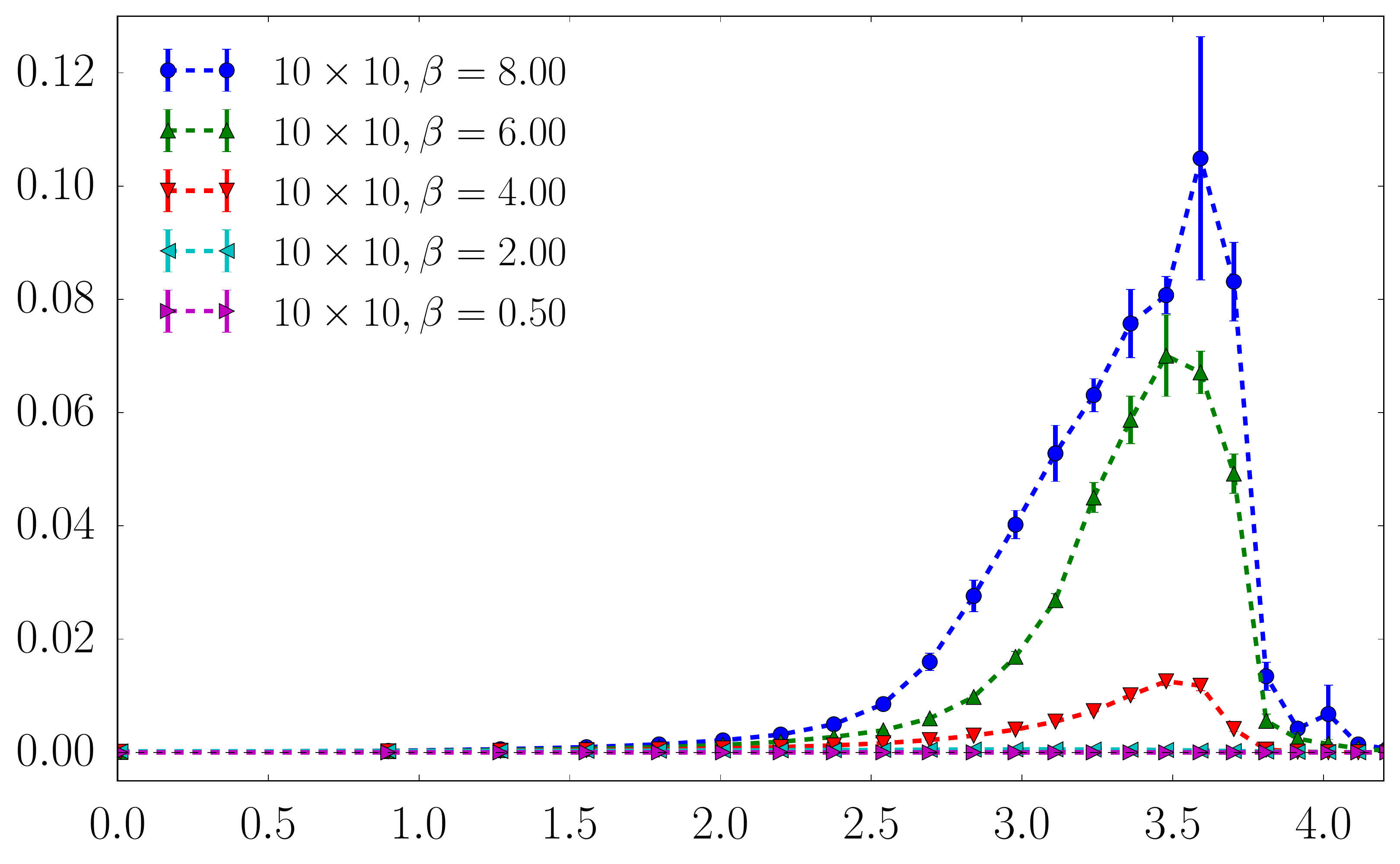}%
		\setlength{\unitlength}{\linewidth}%
		\begin{picture}(0,0)
		\put(-0.5, -0.015){$g$}
		\put(-1.0, 0.22){\rotatebox{90}{$P^\text{s}_{\br = (L/2,L/2)}$}}
		\put(-0.5, 0.605){$g$}
		\put(-1.0, 0.87){\rotatebox{90}{$C_{\tau=0}(\pi,\pi)$}}
		\put(-1.02, 1.17){(a)}
		\put(-1.02, 0.55){(b)}
		\end{picture}
	\end{center}
	\caption{\label{Fig:8} Equal-time nematic correlation function for a $10\times 10$ lattice as a function of $\beta$. The correlation functions are (a) for the AFO order as defined for Fig.~\ref{Fig:3} and (b) the on-site superconductivity as defined for Fig.~\ref{Fig:5}(b). The value increases dramatically in the ordered regions as $\beta$ increases, supporting the presence of order as presented in the main text. However, the small lattice sizes give rise to large finite-size contributions, even when the extrapolation to the thermodynamic limit indicates no order.}
\end{figure}
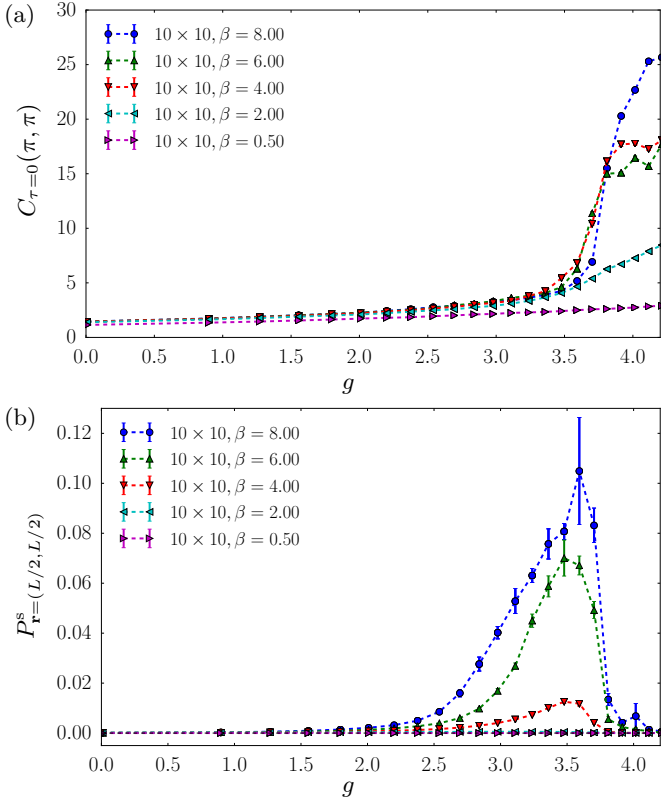

Throughout the main paper, we have shown data at a fixed temperature $\beta = 8$, where the AFO and superconducting orders where well developed on the simulated system sizes and finite-size extrapolation indicated the presence of order. Here in Fig.~\ref{Fig:8}, we illustrate the temperature dependence of the AFO equal-time correlation function~\eqref{Eq:nem}  and of the correlation function for on-site superconducting order~\eqref{Eq:Ps} at maximum separation $\br = (L/2, L/2)$.

Both {AFO and superconducting} correlation functions show an increase as the temperature is lowered, as expected in regions where order appears. Because of large finite-size effects, significant increases are seen even at intermediate temperatures such as $\beta = 2,4$ where finite-size scaling indicates {an absence of long-range order in the thermodynamic limit.}   At temperatures $\beta = 8$, \noteM{extrapolation shows the presence of} AFO and superconducting orders.  \note{At $\beta = 6$, large error bars in the finite-size extrapolations do not allow one to rigorously establish the presence of either order. Tentatively, there is an indication that superconducting order is already established and the superconducting transition is at $\beta < 6$, although simulations on larger lattice sizes are needed to refine this estimate.}

\section{Nematic and AFO Fluctuations} \label{sec:app:b}

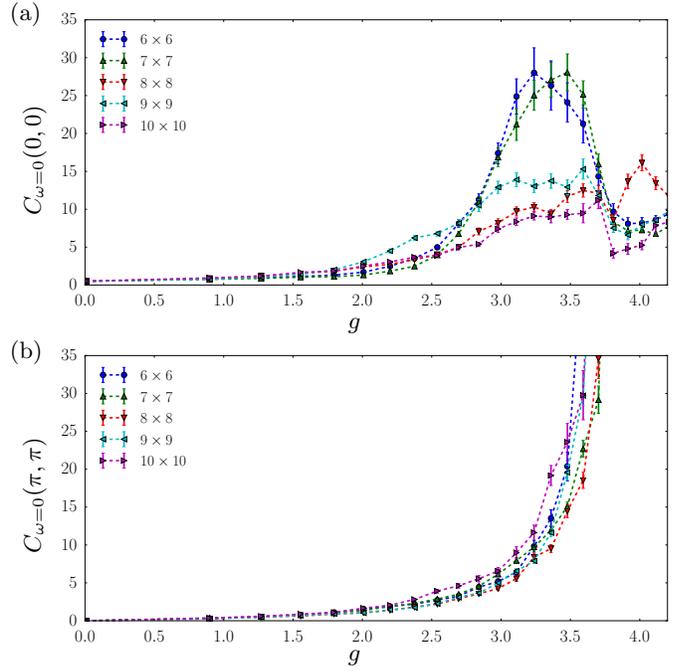
\begin{figure}[tb]
	\begin{center}
		\includegraphics[width=0.95\columnwidth]{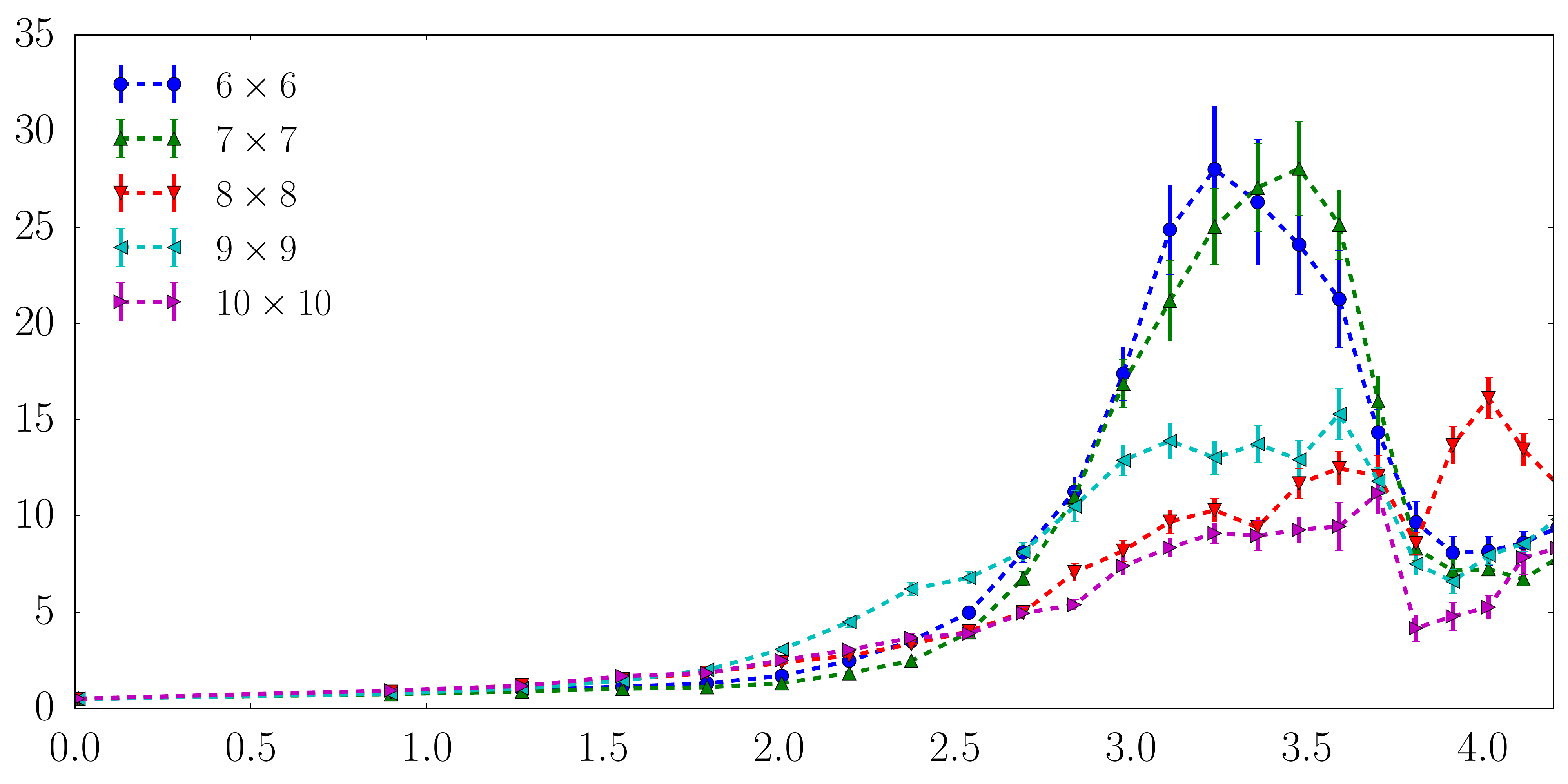}\\
		\vspace{1em}
		\includegraphics[width=0.95\columnwidth]{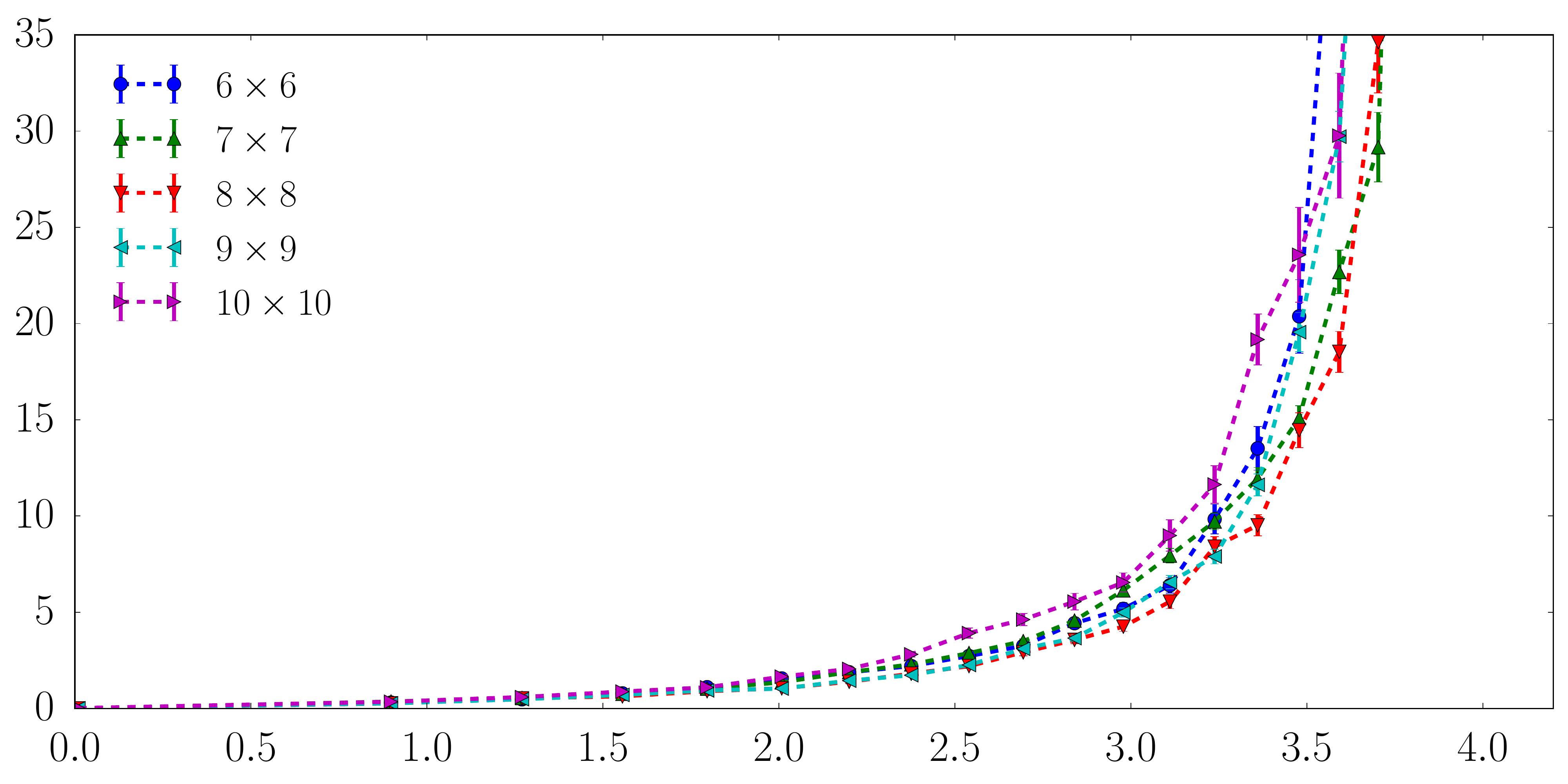}%
		\setlength{\unitlength}{\linewidth}%
		\begin{picture}(0,0)
		\put(-0.5, -0.015){$g$}
		\put(-1.0, 0.15){\rotatebox{90}{$C_{\omega=0}(\pi,\pi)$}}
		\put(-0.5, 0.495){$g$}
		\put(-1.0, 0.67){\rotatebox{90}{$C_{\omega=0}(0,0)$}}
		\put(-1.02, 0.97){(a)}
		\put(-1.02, 0.45){(b)}
		\end{picture}
	\end{center}
	\caption{\label{Fig:9} Susceptibility for (a) the nematic and (b) the AFO order, derived from the connected correlation function of the Hubbard-Stratonovitch fields. The transition to the AFO is at $g  \sim 3.7$; for $g \gtrsim 3.4$ the fluctuations of the AFO dominate as expected close to the transition. For weaker $g$ the nematic fluctuations are stronger for smaller lattice sizes as discussed in the main text.}
\end{figure}

Figure \ref{Fig:9} shows the nematic susceptibility at $\omega = 0$, calculated via the connected correlation function of the Hubbard-Stratonovitch fields

\begin{equation}
C_{\omega}(\bq) =  \frac{1}{L^2 \beta}\sum_{\bi,\bj, \tau}e^{i\bq\cdot (\bi-\bj)- i \tau \omega}\left[\corr{\varphi_{\tau, \bi}\varphi_{0,\bj}} - \corr{\varphi_{\tau, \bi}}\corr{\varphi_{0,\bj}} \right].
\end{equation} 

\noindent The fields $\varphi$ are the conjugate variable to $\delta n$ and therefore this correlation function is directly proportional to the strength of fluctuations. 
The Hubbard-Stratonovitch fields have \note{significantly} slower convergence compared to the fermion Green's functions \note{and should be interpreted with caution}. \note{However,} there are two general aspects which might be gained from the data. First, that the uniform nematic and AFO fluctuations are of the same order of magnitude until very close to the AFO transition, where superconductivity is suppressed. Second, that the peak in the uniform fluctuations on short lattices (``short-range order'') is a not an artifact of the equal-time correlation function of Fig.~\ref{Fig:5}(a), but indeed a signal of enhanced uniform fluctuations.
}

\bibliography{nematic}
\end{document}